\documentclass[12pt]{article}

\usepackage{amssymb,amsthm,amsfonts,epsfig, multicol,mathrsfs,bm}

\def\gappeq{\mathrel{\rlap {\raise.5ex\hbox{$>$}} {\lower.5ex\hbox{$\sim$}}}}
\def\lappeq{\mathrel{\rlap{\raise.5ex\hbox{$<$}} {\lower.5ex\hbox{$\sim$}}}}
\def\beq{\begin{equation}} \def\eeq{\end{equation}}
\def\bea{\begin{eqnarray}} \def\eea{\end{eqnarray}}
\def\bq{\begin{quote}} \def\eq{\end{quote}}
\def\bc{\begin{center}} \def\ec{\end{center}}

\def\nn{\nonumber}
\def\FC{{\rm FC}}
\def\FV{{\rm FV}}
\def\Pl{{\rm Pl}}

\def\meg{\mu \rightarrow e \gamma}\def\tmg{\tau \rightarrow \mu \gamma}
\def\teg{\tau \rightarrow e \gamma}

\begin{document}
\pagestyle{empty}
\begin{flushright}
CERN-PH-TH/2007-023 \\
DFPD-07/TH/01\end{flushright} \vspace*{2cm}

\begin{center}
{\Large\textbf{Observable Electron EDM and Leptogenesis}
}\vspace*{0.80cm}

{\large \sf F.~R. Joaquim $^{a,*}$, I. Masina
$^{b,\S}$ and A. Riotto $^{a,c,\ddag}$}\\[0.4cm]

$^{a}$\textit{Dipartimento di Fisica ``G. Galilei'', Universit\`a di
Padova and INFN, Sezione di Padova, Via Marzolo, 8 - I-35131 Padua - Italy} \\

$^b$ \textit{CERN, Department of Physics, Theory Division, CH-1211
Geneva 23, Switzerland}

$^c$ \textit{D\'epartement de Physique Th\'eorique, Universit\'e de
Gen\`eve, 24 Quai Ansermet, Gen\`eve, Switzerland}
\end{center}

\vskip 1cm
\begin{center}{\bf Abstract} \end{center}

\begin{quote}
{In the context of the minimal supersymmetric seesaw model, the
CP-violating neutrino Yukawa couplings might induce an electron EDM.
The same interactions may also be responsible for the generation of
the observed baryon asymmetry of the Universe via leptogenesis. We
identify in a model-independent way those patterns within the seesaw
models which predict an electron EDM at a level probed by planned
laboratory experiments and show that negative searches on $\teg$
decay may provide the strongest upper bound on the electron EDM. We
also conclude that a possible future detection of the electron EDM
is incompatible with thermal leptogenesis, even when flavour effects
are accounted for.} \end{quote}

\vspace*{0.8cm}

\noindent $^*$ joaquim@pd.infn.it\\
$^\S$ isabella.masina@cern.ch\\
$^\ddag$ riotto@pd.infn.it

\newpage
\setcounter{page}{1}
\pagestyle{plain}
\def\thefootnote{\arabic{footnote}}
\setcounter{footnote}{0}

%%%%%%%%%%%%%%%%%%%%%%%%%%%  INTRODUCTION %%%%%%%%%%%%%%%%%%%%%%%%%%%%%%%%%

\section{Introduction}

New physics has to be invoked in order to explain why neutrinos are
massive and mix among each other, and also why they turn out to be
much lighter than the other known fermions. Among the proposed
explanations, the idea that neutrino mass suppression is due to the
decoupling of heavy states at a very high energy has prevailed in
the last three decades. This principle is behind the formulation of
the well-known seesaw mechanism~\cite{seesaw}.

Besides many experiments aiming to characterize neutrino
oscillations, several other present and future experiments are
planned to search for alternative signals of lepton flavour
violation (LFV), {\em e.g.} in radiative charged lepton decays. It
is therefore of great importance to explore theoretical scenarios
where LFV is enhanced to levels at reach of planned experiments. One
of the most appealing theoretical frameworks where this may actually
occur relies on supersymmetry (SUSY). Within the minimal
supersymmetric extension of the standard model (MSSM), LFV can be
communicated to the SUSY-breaking sector by the same interactions
which participate in the seesaw mechanism producing a misalignment
between leptons and sleptons~\cite{Borzumati:1986qx}. This turns out
to be particularly interesting in the case where the dynamics
responsible for breaking SUSY is flavour-blind, like in
minimal-supergravity scenarios (mSUGRA).

The phenomenological aspects related to supersymmetric seesaw
mechanisms have been widely studied in the literature, especially
those which concern LFV
decays~\cite{rgess,Casas:2001sr,Lavignac:2001vp,Masina:2002mv,
Joaquim:2006uz}. In particular, most of the analysis have been based
on the extension of the MSSM particle content where right-handed
(RH) heavy neutrinos (responsible for the neutrino mass suppression)
are added. RH neutrinos may play a role in the generation of the
baryon asymmetry. If their couplings to the ordinary charged leptons
are complex and therefore violate CP, then their out-of-equilibrium
decays may induce the observed baryon asymmetry of the Universe,
$Y_B=(0.87\pm 0.03)\times 10^{-10}$~\cite{wmap}, via the
leptogenesis mechanism~\cite{Fukugita:1986hr}. In the simplest
thermal scenario, the lightest heavy Majorana neutrino is produced
after inflation by thermal scatterings and subsequently decays
out-of-equilibrium violating CP and lepton number. Flavour effects
play also a role in thermal leptogenesis since its dynamics depends
on which charged lepton Yukawa interactions are in thermal
equilibrium and, therefore, the computation of $Y_B$ depends on
which temperature regime the heavy Majorana neutrino decays occur.
We will come back to this point later.

Besides flavour violation, also CP-violating (CPV) effects may be
induced in the soft SUSY breaking terms by the CPV seesaw
interactions. These may lead to relevant contributions to the
charged lepton electric dipole moments (EDM), in particular that of
the
electron~\cite{Ellis:2001xt,Ellis:2002xg,Masina:2003wt,Farzan:2004qu,Masina:2005am}.
The present upper bound on the electron EDM is $d_e< 1.6\times
10^{-27}$ e cm at $90\%$ C.L.~\cite{Regan:2002ta}. Within three
years, the Yale group plans to reach a sensitivity of about
$10^{-29}$ e cm~\cite{DeMille} and hopefully go down to $10^{-31}$ e
cm within five years. A more ambitious proposal, based on
solid-state physics methods, exists to probe $d_e$ down to
$10^{-35}$ e cm~\cite{Lamoreaux}.

The above discussion shows that, by extending the MSSM particle
content with heavy neutrino singlets, a new window is widely opened
into the investigation of new effects in flavour physics and
cosmology. Ultimately, one expects to be able to relate various
phenomena like neutrino oscillations, LFV, leptogenesis and EDMs.
This programme is of extreme importance and may shed some light over
the origin of neutrino masses and mixing and its relation with
possible new physics observations. In what follows, we consider the
scenario where the (dominant) LFV and CPV effects in the SUSY soft
breaking sector are exclusively generated by the seesaw Yukawa
interactions. Under these assumptions, we identify the Yukawa
structure leading to an observable $d_e$ and at the same time
compatible with the present bounds on LFV decays and show that there
is an upper bound on $d_e$ coming from negative searches of the
$\teg$ decay. This indirect bound can be even stronger than the
direct-search limit. We will also study the impact on thermal
flavoured-leptogenesis for cases where $d_e$ is at hand of future
experiments, showing that a positive detection of $d_e$ is
incompatible with thermal leptogenesis.

The paper is organized as follows. In Section 2 we review the issue
of EDM and LFV within supersymmetric seesaw models and discuss the
bound on $d_e$ provided by the $\teg$. In Section 3 we give a short
review of thermal leptogenesis, including flavour effects. In
Section 4 we discuss the patterns of neutrino Yukawas giving rise to
a large $d_e$, but still compatible with present bounds on $\meg$,
$\teg$ and $\tmg$. The conclusions are drawn in Section 5.

%%%%%%%%%%%%%%%%%%%%%%%%%%%%%%%%%%%%%%%%%%%%%%%%%%%%%%%%%%%%%%%%%%

\section{EDM and LFV from seesaw Yukawa couplings}

We consider the framework of the minimal supersymmetric standard
model (MSSM) extended with three heavy Majorana neutrino singlets
$N$. Working in the basis where the charged lepton mass matrix
$m_{\ell}$ is real and diagonal, the seesaw~\cite{seesaw}
interactions at high energy are described by the following
superpotential terms
\beq
{\cal W_{SS}}= N\, Y_\nu \,L\, H_u +\frac{1}{2} N \hat M N\;\;,\;\;
\hat M={\rm diag}(M_1,M_2,M_3)\,,
\eeq
where $Y_\nu$ is the Dirac neutrino Yukawa coupling matrix and $M_i$
stand for the (positive) heavy Majorana neutrino masses ordered as
$M_1<M_2<M_3$. As usual, $L$ denotes the lepton doublets and $H_u$
the hypercharge $1/2$ Higgs superfield. Integrating out the heavy
neutrino singlets, after electroweak symmetry breaking one obtains
the effective Majorana neutrino mass matrix
\beq
m_\nu^{eff}=Y_\nu^T \frac{1}{\hat M} Y_\nu v_u^2 = U^* \hat m\,
U^\dagger \;\;,\;\; \hat m={\rm diag}(m_1,m_2,m_3)\,,
\eeq
where $U$ is the leptonic mixing matrix (parameterized in the usual
way, see Eq.~(\ref{parU})), $m_i$ are the (positive) effective
neutrino masses, $v_u=v\sin\beta$ with $v=174\,{\rm GeV}$ and
$\tan\beta=\langle H_u\rangle/\langle H_d \rangle$.

Because of RGE running from high to low energy scales, the seesaw
Yukawa couplings potentially induce flavour and CP violations in
slepton masses. In the following we review their impact by assuming
universal and real boundary conditions at the Planck scale $M_{\rm
Pl}$: $m^2_X=m_0^2$ for the soft scalar masses, $A_X=a_0 Y_X$ for
the trilinear $A$-terms and $\tilde M_i=M_{1/2}$ for the gaugino
mass parameters. In addition, we assume that the $\mu$-term is real.
Under these circumstances, the flavour and CP violations in the
low-energy slepton masses arise from those present in the seesaw
Yukawa couplings through RGE running.

The most relevant flavour misalignment is induced by the seesaw
Yukawas in the slepton doublet mass matrix
$m^2_L$~\cite{Borzumati:1986qx}. Following the widely used mass
insertion approximation, the flavour-violating entries of $m^2_L$
can be parametrized by
\beq \delta^{LL}_{ij}  =  \frac{m^2_{L ij}}{\bar m^2_L} = -
\frac{1}{(4 \pi)^2} \frac{6 m_0^2 + 2 a_0^2}{\bar m^2_L}~ C_{ij}
~~~~(i\neq j)~,
\eeq
where $\bar m_L$ is an average doublet slepton mass and $C_{ij}$
encodes the dependence of $\delta^{LL}_{ij} $ on the Yukawa
interactions:
\beq
 \label{Cgen}
C_{ij}  =  \sum_{k} C_{ij}^k~~~~~,~~~~C_{ij}^k = Y^*_{\nu ki} Y_{\nu
kj}^{} \ln \frac{ M_\mathrm{Pl} }{ M_k }\,.
\eeq
For later convenience, we have isolated the contribution of each
$N_k$ to $C_{ij}$ denoting it by $C^k_{ij}$. Notice that in
Eq.~(\ref{Cgen}) $Y_\nu$ has to be evaluated at $M_{\rm Pl}$.
The induced LFV decays are
\beq
\label{BR}
\mathrm{BR}(\ell_i \rightarrow \ell_j \gamma) =
\mathrm{BR}(\ell_i \rightarrow \ell_j \bar \nu_j \nu_i)~F_B ~
\tan^2\beta~ |\delta^{LL}_{ij}|^2 ~~,
\eeq
where $F_B$ is an adimensional function of supersymmetric masses
which includes the contributions from chargino and neutralino
exchange -- see {\em e.g.}~\cite{Masina:2002mv} and references
therein.

From the above considerations, it is clear that the present
experimental limits on LFV decays can be translated into upper
bounds on $|C_{ij}|$, denoted by $C^{\rm ub}_{ij}$, which depend on
$\tan\beta$ and supersymmetric masses~\cite{Lavignac:2001vp}. In
mSUGRA, once the relation between $a_0$ and $m_0$ has been assigned
and radiative electroweak breaking required, the supersymmetric
spectrum can be expressed in terms of two masses, {\em e.g.} the
bino mass $\tilde M_1$ and the average singlet charged slepton mass
$\bar m_R$ at low energy (we recall that $\tilde M_1 \approx 0.4
M_{1/2}$ and $\bar m_R^2 \approx m_0^2 + 0.15 M_{1/2}^2$). For
definiteness, we will take from now on $a_0=m_0+M_{1/2}$.
Considering in particular the point $P=(\tilde M_1,\bar
m_R)=(200,500)\,{\rm GeV}$, for which the SUSY contribution to the
anomalous magnetic moment of the muon $\delta a_\mu$ is within the
observed discrepancy between the SM and experimental result for
$\tan\beta> 35$, we obtain
\beq
C^{{\rm ub}}_{21}\simeq 5\times10^{-3} \frac{50}{\tan\beta}~~~~,~~~~
C^{{\rm ub}}_{32}\simeq 0.8 \,\frac{50}{\tan\beta} ~~~~,~~~~ C^{{\rm
ub}}_{31}\simeq \frac{50}{\tan\beta} ~~.
\eeq
The strongest constraint
comes from $\meg$, although also those from $\teg$ and $\tmg$ become
non-trivial for $\tan\beta \gtrsim 10$, as they imply that $Y_\nu$
couplings are at most of ${\cal O}(1)$. This, in turn, supports the
use of the perturbative approach. In Fig.~\ref{fig1} of the Appendix
we display the dependence of $C^{{\rm ub}}_{ij}$ in the plane
$(\tilde M_1,\bar m_R)$. %for $a_0=m_0+M_{1/2}$ .

There are two potentially important sources of CP violation induced
by the seesaw Yukawa couplings in the doublet slepton mass matrix.
One is associated to the flavour-conserving (FC)
$A$-terms~\cite{Ellis:2001xt, Ellis:2002xg, Masina:2003wt}, while
the other, generically dominant for $\tan\beta \gtrsim 10$, is
mediated by flavour-violating (FV) $\delta$'s~\cite{Masina:2003wt,
Farzan:2004qu}. The corresponding contributions to lepton EDMs are
given by
\beq
\label{di} d^{\FC}_{i} ~[e~{\rm cm}]= F_{d}
~m_{\ell_i}~\mathrm{Im}(a_{i})~~~~~,~~~~~~ d^{\FV}_{i} ~[e~{\rm
cm}]= F''_d ~\mu\tan\beta~ \mathrm{Im}(\delta^{RR} m_{\ell}\,
\delta^{LL} )_{ii}\,,
\eeq
where $F_d$, $F''_d$ (with dimension of mass$^{-2}$) are functions
of the slepton, chargino and neutralino masses -- see
e.g.~\cite{Masina:2002mv} and references therein. These sources of
CPV are
\beq
\mathrm{Im}(a_i)= \frac{8 a_0}{(4 \pi)^4}~I^{\FC}_i ~~~ ,~~~
\mathrm{Im}(\delta^{RR} m_{\ell} \delta^{LL} )_{ii} = \frac{8
m_{\ell_i}}{(4 \pi)^6}  \frac{(6 m_0^2 + 2 a_0^2) (6 m_0^2 + 3
a_0^2)}{\bar m_L^2 \bar m_R^2} \frac{m^2_\tau \tan^2\beta}{v^2}
~I^{\FV}_i ~,
\eeq
\noindent where
\beq
\label{IeFCIeFV}%
I^{\FC}_i = \sum_{k > k'} I^{(k k'){\FC}}_i~~,~~I^{\FV}_i = \sum_{k
> k'} I^{(k k')\FV}_i\,.
\eeq
Adopting a notation which renders more explicit the link with LFV
decays and defining ${\ln}^{a}_{b}=\ln(M_a/M_{b})$ for short, one
has~\cite{Masina:2005am}
\beq
I^{(k k')\FC}_i=\frac{ \ln^k_{k'} }{\ln^{\Pl}_{k'} }~\mathrm{Im} (
C^{k} C^{k'} )_{ii}~~~,~~~ I^{(k k')\FV}_i =
\widetilde{\ln}^{k}_{k'} ~\mathrm{Im} \!\left(\!C^{k} \,\frac{ m^2
_{\ell}}{m^2_\tau} \,C^{k'} \!\right)_{\!ii}\,,%
\label{LeI}
\eeq
with $\widetilde{ \ln }^3_2= \ln^3_2$, $\widetilde{ \ln}^3_1 =
\ln^3_1 (1 - 2 \ln^3_2    \ln^2_1 / \ln^3_1 \ln^{\Pl}_1 )$,
$\widetilde{ \ln }^2_1= \ln^2_1 (1 - 2 \ln^{\Pl}_3 \ln^3_2/
\ln^{\Pl}_2 \ln^{\Pl}_1 )$. Notice that these contributions arise as
an effect of a splitted spectrum of right-handed neutrinos and would
vanish in the case of a degenerate spectrum~\cite{Ellis:2001xt,
RomStru}. It turns out that the seesaw-induced contributions to
$d_\mu$ and $d_\tau$ are below the planned experimental
sensitivities; on the contrary the seesaw-induced contribution to
$d_e$ might be at the level of planned experiments. The present
experimental upper limit, $d_e^{exp}=1.4\times 10^{-27}$ e cm, can
correspondingly be translated into upper bounds on $|I_e^{\FC}|$ and
$|I_e^{\FV}|$, with a dependence on $\tan\beta$ and supersymmetric
masses - see~\cite{Masina:2003wt,Masina:2005am} for more details.
Within the mSugra framework, we display these bounds in
Fig.~\ref{fig1} of the Appendix. In particular, for the point $P$
introduced before, the upper bounds are
\beq
\label{upperb}%
I_e^{{\rm ub},\FC} \simeq 10^3~~~~,~~~~I_e^{{\rm
ub},\FV} \simeq 10^2~ \left(\frac{50}{\tan\beta}\right)^3\,.
\label{Iub}
\eeq
For comparison, it is useful to estimate the upper allowed values of
$I_e^{\FC}$ and $I_e^{\FV}$ assuming perturbativity: by allowing the
relevant Yukawa couplings of $Y_\nu$ to be of ${\cal O}(1)$ and the
logarithms to be large enough, from Eq.~(\ref{LeI}) one obtains
$I_e^{\FC}\lesssim 50$ and $I_e^{\FV} \lesssim 300$. This means that
in point $P$ the FC-type seesaw-induced contribution, $d_e^{\FC}$,
is below the level of the present experimental sensitivity; an
improvement by two orders of magnitude, pushing $d_e^{exp}$ at the
level of $10^{-29}$ e cm, would be needed to start testing it. On
the contrary, the FV-type seesaw-induced contribution, $d_e^{\FV}$,
might have already exceeded the present experimental bound in $P$
for $\tan\beta\gtrsim 35$ (and for even smaller $\tan\beta$ if the
supersymmetric masses are taken to be smaller than those in $P$).

\begin{figure}
\label{fig2}
\begin{center}
\begin{tabular}{c}
\includegraphics[width=9cm]{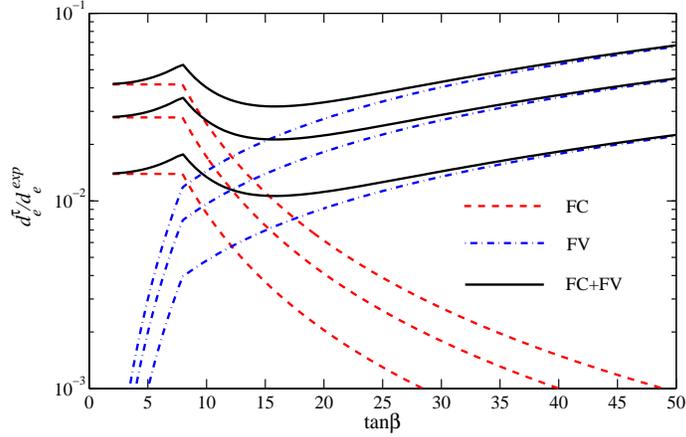}
\end{tabular}
\end{center}
\vspace*{-0.5cm} \caption{Ratios $\min(I_e^{\tau\FC},\ln^3_2
50/\ln^\Pl_2)/I_e^{\rm{ub}\FC}$ (dashed) and
$\min(I_e^{\tau\FV},\ln^3_2 50)/I_e^{\rm{ub}FV}$ (dash-dotted)
evaluated in $P$ as a function of $\tan\beta$. Their sum is
$d_e^\tau/d^{exp}_e$ (solid). We take $M_2=5\times 10^{14}$~GeV and
$M_3/M_2= 10^3,10^2,10$ from top to bottom, respectively.}
\end{figure}

However, as we now turn to discuss, also the experimental limit on
$\teg$ provides an upper bound to the seesaw-induced $I_e^{\FC}$ and
$I_e^{\FV}$. At present, such indirect bounds are even stronger than
those from direct searches of the electron EDM. Once the constraint
from $\teg$ is taken into account, it turns out that both the FC and
FV-type seesaw contributions have to be sizeably smaller than what
would be allowed assuming perturbativity. The argument is based on
the fact that, barring cancelations, $|C^k_{ij}|< C^{{\rm ub}}_{ij}$
for each $k$. In Eq.~(\ref{LeI}), the strong bound from $\meg$ makes
the terms involving $C^k_{12}$ to be negligibly small. If large,
$I_e^{\FC}$ and $I_e^{\FV}$ are then proportional to the same
combination of Yukawas~\cite{Masina:2005am}:
\bea
I_e^{\FC}\approx \sum_{k >
k'}\frac{\ln^{k}_{k'}}{\ln^{\Pl}_{k'}}\,{\rm
Im}(C^k_{13}C^{k'*}_{13})~~~~,~~~ I_e^{\FV}\approx \sum_{k >
k'}\widetilde\ln^k_{k'} ~{\rm Im}(C^k_{13}C^{k'*}_{13})\,.
\eea
In the discussion of the next sections, it will turn out that the
only relevant contribution comes from $(k,k')=(3,2)$. Therefore, in
what follows, we will focus on $I_e^{(32)\FC} \ln^{\Pl}_{2} \approx
I_e^{(32)\FV}\approx \ln^3_{2} {\rm Im}(C^3_{13}C^{2*}_{13}) $.
Requiring ${\rm Im}(C^k_{13}C^{k'*}_{13})\le{C^{\rm ub}_{13}}^2$,
one thus obtains an indirect upper bound from $\teg$, to be denoted
by $I_e^\tau$. Considering in particular the point $P$, the indirect
upper bounds from $\teg$ are
\beq
I_e^{\tau \FC} \ln^{\Pl}_{2}=I_e^{\tau \FV} =  \ln^3_2
\left(\frac{50}{\tan\beta}\right)^2\,.
\eeq
As a comparison, allowing the relevant Yukawa couplings $Y_\nu$ to
be (at most) of ${\cal O}(1)$ and the logarithms involved to be large,
%again $(M_2,M_3)\simeq(10^{-5},10^{-2})\,M_{\Pl}$,
one would have ${\rm Im}(C^3_{13}C^{2*}_{13}) \lesssim 50$. This means that, for
$\tan\beta\gtrsim 5$, both $I_e^{\tau \FC}$ and $I_e^{\tau \FV}$ are
smaller than what allowed by perturbativity. Typically,  $10\,
I_e^{\tau \FC}\sim I_e^{\tau \FV}\sim 500$ for $\tan \beta= 5$,
while $10\, I_e^{\tau \FC} \sim I_e^{\tau \FV}\sim 5$ for $\tan
\beta= 50$. The indirect bounds from $\teg$ are thus stronger than
those from direct searches of $d_e$, shown in Eq.~(\ref{Iub}).
Taking $P$ and representative splittings between $M_3$ and $M_2$, we
display in Fig.~\ref{fig2} the ratio $d_e^\tau/d_e^{exp}= I_e^{\tau
\FC}/I_e^{{\rm ub},\FC}+I_e^{\tau \FV}/I_e^{{\rm ub},\FV}$ as a
function of $\tan\beta$. It turns out that the indirect upper bound
from $\teg$, $d_e^\tau$, is stronger than the present direct one,
$d_e^{\rm exp}$, by about two orders of magnitude\footnote{ As a
consequence, a potential discovery of the electron EDM within an
order of magnitude from the present sensitivity should not to be
interpreted as due to the seesaw-induced effects - this of course
holds for mSugra and point $P$.}. Planned experiments lowering the
bounds on $d_e$ by about three orders of magnitude will then provide
sensible tests of the seesaw-induced effects. Clearly they will test
the seesaw models that maximize $I_e^{\FV}$ (hence also
$I_e^{\FC}$), {\em i.e.} the models where: 1) $M_3\gtrsim 10~ M_2$;
2) ${\rm Im}(C^3_{13}C^{2*}_{13}) \sim {C^{\rm ub}_{13}}^2$. The
latter of course also implies that $\teg$ is close to the
experimental bound.

%%%%%%%%%%%%%%%%%%%%%%%%%%%%%%%%%%%%%%%%%%%%%%%%%%%%%%%%%%%%%%%%%%%
\section{Thermal leptogenesis including flavours}

\noindent In this section we provide a short review of leptogenesis
which will be useful to establish the link between the high energy
CP-violation responsible for the generation of the baryon asymmetry
through thermal leptogenesis and the one responsible for the EDM of
the electron.

Thermal leptogenesis~\cite{lept,ogen,work} takes place through the
decay of the lightest of the heavy Majorana neutrinos if these are
present in the early Universe. The out-of-equilibrium decays occur
violating lepton number and CP, thus satisfying the Sakharov's
conditions~\cite{sakharov}. In grand unified theories (GUT) the
heavy Majorana neutrino masses are typically smaller than the scale
of unification of the electroweak and strong interactions, $M_{\rm
GUT} \simeq 2\times 10^{16}$~GeV, by a few to several orders of
magnitude. This range coincides with the range of values of the
heavy Majorana neutrino masses required for a successful thermal
leptogenesis.

We will account for the flavour effects which appear for $M_1\lappeq
10^{12}$~GeV and have been recently investigated in  detail
in~\cite{Barbieri99,endoh,davidsonetal,nardietal,dibari,
davidsonetal2,antusch,silvia1,Branco:2006ce,aat,silvia2,adsar}
including the quantum oscillations/correlations of the asymmetries
in lepton flavour space~\cite{davidsonetal,adsar}. The Boltzmann
equations describing the asymmetries in flavour space have
additional terms which can significantly affect the result for the
final baryon asymmetry. The ultimate reason is that realistic
leptogenesis is a dynamical process, involving  the  production and
destruction of the heavy RH neutrinos, and of a lepton asymmetry
that is distributed among {\it distinguishable} lepton flavours.
Contrarily to what is generically assumed in the one-single flavour
approximation, the $\Delta L=1$ inverse decay processes which
wash-out the net lepton number are flavour dependent, that is the
lepton asymmetry carried by, say, electrons can be washed out only
by the inverse decays involving the electron flavour. The
asymmetries in each lepton flavour, are therefore washed out
differently, and will appear with different weights in the final
formula for the baryon asymmetry. This is physically inequivalent to
the treatment of wash-out in the one-flavour approximation, where
the flavours are taken indistinguishable, thus obtaining  the
unphysical result that, {\em e.g.}, an asymmetry stored in the
electron lepton charge may be washed out by inverse decays involving
the muon or the tau charges.

When flavour effects are accounted for, the final value of the
baryon asymmetry is the sum of three contributions. Each term is
given by the CP asymmetry in a given lepton flavour $\ell$, properly
weighted by a wash-out factor induced by the same  lepton number
violating processes. The wash-out factors are also flavour
dependent.

Of course, since we are dealing with the MSSM, we have  to consider
the presence of the supersymmetric partners of the RH heavy
neutrinos, the sneutrinos $\widetilde{N}_i$ ($i=1,2,3)$, which  also
give a contribution to the flavour asymmetries, and of the
supersymmetric partners of the lepton doublets, the slepton
doublets. Since the effects of supersymmetry breaking may be safely
neglected, the  flavour CP asymmetries in the MSSM are twice those
in the SM and double is also the possible channels by which a lepton
flavour asymmetry is reproduced. However, the $\Delta L=1$
scatterings washing out the asymmetries are also doubled and the
number of relativistic degrees of freedom is almost twice the  one
for  the SM case. As a result, introducing new degrees of freedom
and interactions does not appreciably change the flavour
asymmetries. There are  however  two other and important differences
with respect to the nonsupersymmetric thermal leptogenesis SM case.
First, in the MSSM, the flavour-independent formulae can only be
applied for temperatures larger than $(1+\tan^2
\beta)\times 10^{12}$ GeV. Indeed, the squared charged lepton Yukawa
couplings in the MSSM are multiplied by $(1+\tan^2 \beta)$.
Consequently, the $\mu$ and $\tau$ lepton Yukawa couplings are in
thermal equilibrium for
 $(1+\tan^2 \beta)\times 10^5 \: \mbox{GeV}
\ll T \ll (1+\tan^2 \beta)\times 10^{9} \: \mbox{GeV}$ and all
flavours in the Boltzmann equations are to be treated separately.
For $(1+\tan^2 \beta)\times 10^9 \: \mbox{GeV} \ll T \ll (1+\tan^2
\beta) \times 10^{12} \: \mbox{GeV}$, only the $\tau$ Yukawa
coupling is in equilibrium and only the $\tau$ flavour is treated
separately in the Boltzmann equations, while the $e$ and $\mu$
flavours are indistinguishable. Secondly, the relation between the
baryon asymmetry $Y_B$ and the lepton flavour asymmetries has to be
modified to account for the presence of two Higgs fields. Between
$(1+\tan^2 \beta)\times 10^5$ and $(1+\tan^2 \beta)\times 10^{9}$
GeV, the relation is
%%%%%%%%%%%%%%%%%%%%
\begin{equation}
Y_{B} \simeq  -\frac{10}{31 g_*} \left[ \epsilon_e \eta
\left(\frac{93}{110} \widetilde{m}_e\right) + \epsilon_\mu \eta
\left(\frac{19}{30} \widetilde{m}_\mu\right) + \epsilon_\tau \eta
\left(\frac{19}{30} \widetilde{m}_\tau \right) \right], \label{a}
\end{equation}
where the flavour lepton asymmetries are computed including
leptons and sleptons.
Between
$(1+\tan^2 \beta)\times 10^9$ and $(1+\tan^2 \beta)\times 10^{12}$ GeV,
the relation is
%%%%%%%%%%%%%%%%%%%%
\begin{equation}
Y_{B} \simeq  -\frac{10}{31 g_*} \left[\epsilon_2 \eta
\left(\frac{541}{761}\widetilde{m}_2\right) + \epsilon_\tau \eta
\left(\frac{494}{761}\widetilde{m}_\tau\right) \right]\,, \label{aa}
\end{equation}
where the number of relativistic degrees of freedom is counted by
$g_*=228.75$. The observed value is $Y_B=(8.7\pm
0.3)\times10^{-11}$~\cite{wmap}. Let us explain the various terms
entering in Eqs.~(\ref{a}) and (\ref{aa}). The CP-asymmetry in each
flavour is given
by~\cite{Barbieri99,nardietal,davidsonetal,davidsonetal2}
\begin{equation}
\epsilon_{\ell} = -\frac{3 M_1}{8\pi v_u^2}\, \frac{{\rm Im}\left(
\sum_{\beta\rho} m_\beta^{1/2}m_\rho^{3/2} U^*_{\ell \beta}U_{\ell
\rho} R_{1\beta}R_{1\rho}\right)}{\sum_\beta
m_\beta\left|R_{1\beta}\right|^2}\,,~~ \ell=e,\mu,\tau\,,
\label{epsa1}
\end{equation}
where the (in general complex) orthogonal matrix $R$~\cite{Casas:2001sr}
is defined as
\beq
\label{CIb}%
Y_\nu  =\frac{\sqrt{\hat M}}{v_u} R\,\sqrt{\hat m} \,U^\dagger\,.
\eeq
\\
Similarly, one defines a ``wash-out mass parameter'' for each
flavour
$\ell$~\cite{Barbieri99,nardietal,davidsonetal,davidsonetal2}:
%%%%%%%%%%%%%%%%%%%%%%%%%%%%%%%%%%%%%%%%%%%%%
\begin{equation}
\left( \frac{\widetilde{m}_\ell}{2\,\sin^2\beta\times 10^{-3}\,{\rm
eV}}\right) \equiv \frac{\Gamma(N_1\rightarrow H\,
\ell)}{H(M_1)}\;\;,\;\; \widetilde{m}_\ell \equiv \frac{|Y_{\nu 1
\ell}|^2\,v_u^2}{M_{1}}= \left|\sum_{k}R_{1k}m_k^{1/2}U_{\ell
k}^*\right|^2 \,. \label{tildmal1}
\end{equation}
\\
The quantity $\widetilde{m}_\ell$ parametrizes the decay rate of
$N_1$ to the leptons of flavour $\ell$. Furthermore, in
Eq.~(\ref{aa}), $\epsilon_2 =\epsilon_{e}+\epsilon_{\mu}$ and
$\widetilde{m}_2=\widetilde{m}_e+ \widetilde{m}_\mu$. Finally,
\begin{equation}
\eta\left(\widetilde{m}_\ell\right)\simeq \left[
\left(\frac{\widetilde{m}_\ell}{8.25\times 10^{-3}\,{\rm
eV}}\right)^{-1}+ \left(\frac{0.2\times 10^{-3}\,{\rm
eV}}{\widetilde{m}_\ell} \right)^{-1.16}\ \right]^{-1}\,
\label{eta1}
\end{equation}
parametrizes the wash-out suppression of the asymmetry due to
$\Delta L=1$ inverse decays and scatterings. Notice that the
wash-out masses $\widetilde{m}_2$ and $\widetilde{m}_\tau$ in
Eq.~(\ref{a}) are multiplied by some numerical coefficients which
account for the dynamics involving the lepton doublet asymmetries
and the asymmetries stored in the charges
$\Delta_\ell=(1/3)B-L_\ell$~\cite{davidsonetal2}.

%%%%%%%%%%%%%%%%%%%%%%%%%%%%%%%%%%%%%%

\section{Maximizing $\bm{d_e}$ with the constraint from $\bm{\meg}$
and impact on thermal leptogenesis}

In this section our goal is to identify those patterns within the
seesaw models which predict $d_e$ at hand of future experiments,
while keeping the prediction for LFV decays, in particular $\meg$,
below the present bound. In this respect, the analysis we perform is
model-independent. From the discussion of the previous section, it
is clear that to have $d_e$ at hand of future experiments, the
quantities $I_e$ defined in Eq.~(\ref{IeFCIeFV}) have to be
maximized. Our aim is also to understand whether these models can
explain the observed baryon asymmetry of the universe via $N_1$
decays. Our work differs from that of Ref.~\cite{Ellis:2002xg} in
two respects. First, our analysis includes also the FV contribution
to the electron EDM, which is dominant for large values of
$\tan\beta$ and, secondly, we include flavour effects in computing
the baryon asymmetry from leptogenesis.

We choose to work with not too heavy RH neutrino $N_1$, say $M_1$ in
the range $(10^{10}-10^{11})$ GeV. This choice has the advantage
that, barring accidental cancelations, the couplings of $N_1$ are
rather small, $Y_{\nu1j} \lesssim x = {\cal O}(10^{-2})$. This has
important consequences: i) $C^1_{21}$ does not exceed the $\meg$
bound; ii) the eventual contributions to $d_e$ from
$I_e^{(31),(21)}$ (both FV and FC) are much smaller than the limit
inferred from experiment and can be neglected in first
approximation. The only potential source of a large $d_e$ is
therefore
\beq
\label{FVFCap}%
I_e^{(32)\FC} \ln^{\Pl}_2 \approx I_e^{(32)\FV} \approx I_4
\ln^{\Pl}_3\ln^{\Pl}_2\ln^3_2~~~,~~~~I_4={\rm Im}( Y_{\nu31}^*
Y_{\nu33}^{} Y_{\nu21}^{} Y_{\nu23}^* )\,, \label{4Y}
\eeq
which is maximized requiring that:\\ %
\noindent 1) the splitting between $M_3$ and $M_2$ is large; \\
\noindent 2) the four Yukawas appearing in Eq.~(\ref{4Y}) are of
${\cal O}(1)$, but satisfy the bound from $\teg$.

Barring conspiracies between $C^2_{21}$ and $C^3_{21}$, to suppress
$\meg$ below the present bound one must require both to be smaller
than $C_{21}^{\rm{ub}}$. Since $C^k_{21} \propto Y_{\nu k2}^* Y_{\nu
k1}^{}$, in order to keep $I_e^{(32)}$ large, one has to impose
$|Y_{\nu32}|, |Y_{\nu33}| \lesssim y \approx
C_{21}^{\rm{ub}}/5$. Explicitly, we are looking for textures of the
form
\beq
Y_\nu = \left( \matrix{%
\lesssim x & \lesssim x & \lesssim x \cr %
{\cal O}(1)&\lesssim y & {\cal O}(1)\cr %
{\cal O}(1) &\lesssim y  & {\cal O}(1)} \right)\,,
\label{Ynu}%
\eeq
where, as already mentioned, $x ={\cal O}(10^{-2})$ ensures that
$C_{21}^1$ satisfies the upper bound on $\meg$. Notice also that
this pattern enhances $\teg$ while suppressing $\tmg$.

It is convenient to exploit the complex orthogonal matrix
$R$~\cite{Casas:2001sr} introduced in Eq. (\ref{CIb}) and adopt a
standard parameterization for the MNS mixing matrix
\beq
U =O_{23}(\theta_{23})\,\Gamma_\delta \,O_{13}(\theta_{13})
\,\Gamma^*_\delta \,O_{12}(\theta_{12}) \times
\textrm{diag}(e^{i\alpha_1/2}, e^{i\alpha_2/2}, 1)\,, \label{parU}
\eeq
where $\Gamma_\delta=$ diag$(1,1,e^{i\delta})$,
$O_{ij}=[(c_{ij},s_{ij})(-s_{ij},c_{ij})]$ with $c_{ij}=\cos
\theta_{ij}$ and $s_{ij}=\sin \theta_{ij}$. The Dirac and Majorana
type phases were denoted by $\delta$ and $\alpha_{1,2}$,
respectively. We adopt for $R$ a parameterization in terms of three
complex angles $\theta^R_{ij}$:
\beq
R=O_{12}(\theta^R_{12})
\,O_{13}(\theta^R_{13})\,O_{23}(\theta^R_{23})
=\left( \matrix{%
c^R_{12} c^R_{13} &  s^R_{12} c^R_{23}-c^R_{12} s^R_{23} s^R_{13} &
s^R_{12} s^R_{23}+c^R_{12} c^R_{23} s^R_{13} \cr %
-s^R_{12}c^R_{13} &  c^R_{12} c^R_{23}+s^R_{12} s^R_{23} s^R_{13}&
c^R_{12} s^R_{23}-s^R_{12} c^R_{23} s^R_{13} \cr
- s^R_{13}        & -s^R_{23} c^R_{13}&c^R_{23}c^R_{13} } \right)\,.
\eeq

As already discussed, the $\meg$ constraint is implemented by
requiring $|Y_{\nu32}|, |Y_{\nu22}| \lesssim y$. In the  approximation
$y=0$ (which turns out to be very satisfactory),
this means that $R$ must satisfy respectively the conditions
\beq
 R_{i3} \sqrt{m_3} U^*_{23} +   R_{i2} \sqrt{m_2} U^*_{22}+
 R_{i1} \sqrt{m_1} U^*_{21} =0~~~~~~,~~~~ i=3,2~~.
\label{la1}
\eeq
Simultaneously we want to enhance
\beq
\label{I42}%
I_4= \frac{M_3 M_2}{v_u^4}{\rm Im}\left[(R \sqrt{\hat
m}\, U^\dagger)^*_{31}(R \sqrt{\hat m}\, U^\dagger)_{33}(R
\sqrt{\hat m} \,U^\dagger)_{21}(R \sqrt{\hat m}
\,U^\dagger)^*_{23}\right].
\eeq
A potentially interesting contribution to the electron EDM requires
$M_3$ and $M_2$ to be as large as possible and, as already stressed,
enough splitted to ensure a large $\ln^3_2$. This also explains why
we cannot rely here on resonant leptogenesis~\cite{resonant}. One
could have, for instance, $M_1 \simeq M_2 \ll M_3$. But since $M_1$
has to be small (for $\meg$), this would imply small $M_2$ and
consequently a suppressed $I_4$. Instead, the pattern $M_1 \ll
M_2\simeq M_3$ implies $\ln^3_2 \ll 1$. Moreover, any baryon
asymmetry resulting from the resonant decays of $N_2$ and $N_3$ can
be washed out by $N_1$ (see however~\cite{vives} and
especially~\cite{Engelhard:2006yg}). Finally, the situation $M_1
\simeq M_2 \simeq M_3$ would lead to a very suppressed
$d_e$~\cite{Ellis:2001xt, RomStru}.

As we are going to discuss in the next section, the structure of
$Y_\nu$ in Eq.~(\ref{Ynu}) determines, to some extent, the form of
$R$. From a qualitative point of view, one can already put forward
some guesses. The requirement of a large atmospheric angle implies
that for some $N_j$ it should be $Y_{\nu j2}\approx Y_{\nu j3}$, but
from Eq.~(\ref{Ynu}), this can happen only for $j=1$. In the case of
hierarchical light neutrinos, this implies dominance of $m_3$ by
$M_1$, while $M_3$ and $M_2$ will be mainly associated to the
lighter masses. Since we aim at $M_2<<M_3$, this means that $M_2$
and $M_3$ will dominate respectively $m_2$ and $m_1$.  According to
the interpretation of $R$ as a dominance
matrix~\cite{Lavignac:2002gf}, these considerations in particular
determine the first row of $R$, which is relevant for leptogenesis,
to be approximately $(0, \sqrt{m_2/m_3}, 1-{\cal O}(m_2/m_3))$.

%%%%%%%%%%%%%%%%%%%%%%%%%%%%%%%%%%%%%%%%%%%%%%%%%%%%%%%%%%%%

In the following we treat separately the
cases of normal and inverted hierarchy for light neutrinos.

\subsection{Normal hierarchy for light neutrinos}

For normal hierarchy (NH) we take $m_3\approx m_@=\sqrt{m_3^2-m_2^2}$, $m_2\approx
m_\odot=\sqrt{m_2^2-m_1^2}$ and leave $m_1\ll m_2$ undetermined. It
is convenient to rewrite the conditions (\ref{la1}) assuming
$R_{i3}\neq 0$,
\beq
-\frac{R_{i2}}{R_{i3}}= \bar t + \frac{R_{i1}}{R_{i3}} \tilde t~\,,
\label{eqRHI}
\eeq
where the quantities $\bar t$ and $\widetilde t$ depend only on the neutrino
parameters at low energy,
\bea
\bar t  &=&\sqrt{\frac{m_3}{m_2}} ~\frac{U^*_{23}}{U^*_{22}}
        = \sqrt{\frac{m_3}{m_2}}\, \frac{e^{i\frac{\alpha_2}{2}}
        t_{23}}{c_{12}} ~\left[1+{\cal O}(s_{13})\right]\,,\nn \\\label{tt}\\
\widetilde t&=&\sqrt{\frac{m_1}{m_2}} ~\frac{U^*_{21}}{U^*_{22}}
        =- \sqrt{\frac{m_1}{m_2}} \,t_{12}\, e^{i \frac{\alpha_2-\alpha_1}
        {2}} ~\left[1+{\cal O}(s_{13})\right]\,\,,    \nn
\eea
with $t_{ij}=\tan\theta_{ij}$. In particular, for a light neutrino spectrum
with NH and taking into
account the present neutrino oscillation data, $|\bar t|\approx 3$
while $|\widetilde t|\ll 1$.

Using the above expressions and introducing the reference scale
$M_@=v_u^2/m_@\approx 5\times 10^{14}$~GeV for the right-handed
neutrino masses, the Yukawa couplings of the second and third rows
of $Y_\nu$ are given by
\begin{eqnarray}
Y_{\nu2i}&\!\!=\!\!&\sqrt{\frac{M_2}{M_@}} \left(\!R_{23}  {\mathbb
U_i^*}
+R_{21}  \sqrt{\frac{m_1}{m_3}} \mathbb{W}^*_i \right)\,,\nn\\
Y_{\nu3i}&\!\!=\!\!&\sqrt{\frac{M_3}{M_@}} \left(\!R_{33}  {\mathbb
U_i^*} +R_{31} \sqrt{\frac{m_1}{m_3}} \mathbb{W}^*_i \right)\,,
\label{eqYHI}
\end{eqnarray}
with
\beq
\mathbb{U}_i = U_{i3} -\frac{U_{23}}{U_{22}}
U_{i2}~~~,~~~~\mathbb{W}_i = U_{i1} - \frac{U_{21}}{U_{22}}\,
U_{i2}\,.
\eeq
Notice that $\mathbb{U}_2$ and $\mathbb{W}_2$ correctly vanish, as
required by the conditions $Y_{\nu 22},Y_{\nu 32}=0$. We also anticipate
that, for $m_1=0$, one has $I_4 \propto {\rm Im}(|{\mathbb U_1}|^2|{\mathbb U_3}|^2)=0$ which,
in the present framework, leads to a very suppressed $d_e$.
For the sake
of the following discussion, we report the approximate expressions
for $\mathbb{U}_{1,3}$ and $\mathbb{W}_{1,3}$ at first order in
$s_{13}$:
\bea
\mathbb{U}_1 &\!\!\!\!=\!\!\!\!& -t_{23} t_{12} (1+t_{23}t_{12}
s_{13} e^{i\delta}) +s_{13} e^{-i\delta} ~~~~,~~~~~
\mathbb{W}_1  =  e^{i\frac{\alpha_1}{2}} \frac{1}{c_{12}}
(1+ s_{13} e^{i\delta} t_{23} t_{12})\,,\nn\\
\mathbb{U}_3 &\!\!\!\!=\!\!\!\!&  \frac{1}{c_{23}} (1+ s_{13}
e^{i\delta} t_{23}t_{12}) ~~~~~~~~~~~~~~~~~~~~~,~~~~~~~~~~~
\!\!\!\!\!\!\!\!\!\!\!\mathbb{W}_3 =   - s_{13}
e^{i(\delta+\frac{\alpha_1}{2})} \frac{1}{c_{23}c_{12}}~~.
\eea
As an example, for tri-bimaximal mixing one has:
$\mathbb{U}_3=\sqrt{2}$, $\mathbb{U}_1=-1/\sqrt{2}$,
$\mathbb{W}_3=0$ and $\mathbb{W}_1=\sqrt{3/2} \,e^{i\alpha_1/2}$.

From the above expressions, and taking into account the smallness of
$\mathbb{W}_3={\cal O}(U_{e3})$, it turns out that large $Y_{\nu23}$
requires $R_{23}={\cal O}(1)$ and $M_2 ={\cal O}( M_@)$. Then, the
condition of large splitting between $M_2$ and $M_3$ implies
$M_3>>M_@$ while large $Y_{\nu 33}$ requires $R_{33}={\cal O}(
\sqrt{M_@/M_3}) < 1$. Explicitly, $R_{33}=c^R_{23} c^R_{13}$ but,
since the condition for $i=3$ in Eq.~(\ref{eqRHI}), $t^R_{23}= \bar
t - {t^R_{13}}\, \widetilde t/ {c^R_{23}}$, naturally suggests
$c^R_{23}$ to be large\footnote{Clearly, $c^R_{23}$ is small only
when $\theta_{23}\approx \pi/2$, but in this case the condition for
$i=3$ in Eq.~(\ref{eqRHI}) forces $c^R_{13}$ to be still small,
$c^R_{13}\approx -\widetilde t$.}, the parameter to be suppressed is
rather $c^R_{13}$. We then define $c^R_{13} =\chi  <1$ and expand at
first order in $\chi$. In particular, from Eq.(\ref{eqYHI}), we have
$c^R_{23}\chi \approx \sqrt{M_@/M_3} Y_{\nu33}/\mathbb{U}_3^*$. The
conditions in Eq.~(\ref{eqRHI}) get now simplified:
\beq
\frac{c^R_{12} c^R_{23} + s^R_{12} s^R_{23}}{-c^R_{12} s^R_{23}
 + s^R_{12} c^R_{23}}+{\cal O}\left(\chi\widetilde t\,\right)
= \bar t = t^R_{23} +   \frac{1}{\chi \,c^R_{23}} \widetilde t\,.
\label{simplyHI}
\eeq
One can then envisage two relevant cases according to the value of
$\theta^R_{12}\,$:

\vskip .5cm
\begin{itemize}

\item If $s^R_{12}=\epsilon <1$, Eq.~(\ref{simplyHI}) gives
$t^R_{23}\approx -1/\bar t +\epsilon$ and $\widetilde t \approx
(\bar t+1/\bar t)\, \chi c^R_{23}$. The latter can be rewritten as
\beq %
\sqrt{\frac{m_1}{m_3}} \sqrt{\frac{M_3}{M_@}} \approx
\frac{s_{23}}{s_{12}} |Y_{\nu33}| ={\cal O}(1)\,.
\eeq
Here, $m_1$ cannot be arbitrarily small, otherwise $M_3$ would
exceed the GUT or Planck scale. At first order in $\epsilon$ one has
$s^R_{23}\approx-\bar c+\epsilon \bar s$, $c^R_{23}\approx\bar
s+\epsilon \bar c$, so that
\beq
R = \left( \matrix{ \chi & \bar c  & \bar s \cr
     0  &  \bar s & -\bar c \cr
     -1 & \chi \bar c & \chi \bar s } \right) +{\cal O}(\epsilon^2, \epsilon \chi,\chi^2)\,.
\eeq
As expected, since $\bar c\approx 1/\bar t<1$ and $\bar s\approx
1-1/(2 \,\bar t^2)$, this structure means that $m_3\approx m_@$ is
dominated by the lightest right-handed neutrino $N_1$, while
$m_2\approx m_\odot$ by $N_2$. Instead, $m_1$ is dominated by the
heaviest, $N_3$, which decouples the more $\chi$ is small. Notice
that such a precise determination of the first row of $R$ allows to
predict leptogenesis, as we are going to discuss. The Yukawa
couplings relevant for $d_e$ are
\beq
Y_{\nu3i}=\sqrt{\frac{M_3}{M_@}} \left(\bar s \chi \mathbb{U}^*_i -
\sqrt{\frac{m_1}{m_3}}  \mathbb{W}^*_i \right)~,~Y_{\nu2i}=
\sqrt{\frac{M_2}{M_@}} \left(-\bar c\, \mathbb{U}^*_i + {\cal
O}(\chi) \sqrt{\frac{m_1}{m_3}} \,\mathbb{W}^*_i \right)\,,
\eeq
so that
\bea
I_4 &=&  \frac{M_2}{M_@} \frac{M_3}{M_@} |\bar c\,|^2
\sqrt{\frac{m_1}{m_3}} \, {\rm Im}\!\left( -\bar s \chi \mathbb{W}_1
\mathbb{U}_1^* |\mathbb{U}_3|^2+\bar s \chi\mathbb{W}_3
\mathbb{U}_3^* |\mathbb{U}_1|^2+ \sqrt{\frac{m_1}{m_3}}
\mathbb{W}_3^* \mathbb{W}_1 \mathbb{U}_1^* \mathbb{U}_3 \right)\nn\\
&=&|Y_{\nu33} Y_{\nu 31} Y_{\nu21} Y_{\nu23}| \left[  -
\frac{m_2}{m_3} \frac{1}{t^2_{23}} \sin\alpha_2 + {\cal
O}(s_{13}\sin\delta) \right]\,.
\eea
The first term is the only one present in the limit $|U_{e3}|=0$ and
displays a suppression by a factor $m_\odot/m_@\simeq 0.17$. For
large values of $|U_{e3}|$ and $\delta \sim \pm \pi/2$ (large
Dirac-type CP violation at low-energies), the second and third terms
can be dominant.

\item If $c^R_{12}=\epsilon <1$, Eq.~(\ref{simplyHI}) gives
$t^R_{23} \approx \bar t\,(1-\epsilon \,\bar t\,)$ and $\widetilde t
\approx  \epsilon \,\bar t \,(\bar t+1/\bar t\,) \chi c^R_{23}$. The
latter can be rewritten as
\beq
\sqrt{\frac{m_1}{m_3}}   \sqrt{\frac{M_3}{M_@}} \approx |\epsilon\,
\bar t\,| ~\frac{s_{23}}{s_{12}}  |Y_{\nu33}| ={\cal O}(\epsilon)\,.
\eeq
In this case $m_1$ is allowed to be small and the limit
$m_1\rightarrow 0$ can be applied. At first order in $\epsilon$ one
has $s^R_{23}\approx\bar s-\epsilon \bar c$, $c^R_{23}\approx\bar
c+\epsilon \bar s$, leading to
\beq
R= \left( \matrix{ 0 &   \bar c  &   \bar s\cr -\chi &  \bar s &  -
\bar c \cr -1  & -\chi \bar s  & \chi \bar c } \right) +{\cal O}
(\epsilon^2,\epsilon\chi,\chi^2)\,.
\eeq
Again, dominance of $N_1$ and $N_2$ has been obtained respectively
for $m_@$ and $m_\odot$, while $m_1$ is associated to $N_3$, which
decouples the more $\chi$ is small. The Yukawa couplings are now
\beq
Y_{\nu3i}=\sqrt{\frac{M_3}{M_@}} \left(\bar c \chi \mathbb{U}^*_i -
\sqrt{\frac{m_1}{m_3}}  \mathbb{W}^*_i \right)~~,~~ Y_{\nu2i}=
\sqrt{\frac{M_2}{M_@}} \left(-\bar c\, \mathbb{U}^*_i+{\cal O}
(\chi) \sqrt{\frac{m_1}{m_3}}  \mathbb{W}^*_i \right)\,,
\eeq
and, consequently,
\bea
I_4 &=& \frac{M_2}{M_@} \frac{M_3}{M_@} |\bar c|^2
\sqrt{\frac{m_1}{m_3}}\, {\rm Im}\left( -\bar c \chi \mathbb{W}_1
\mathbb{U}_1^* |\mathbb{U}_3|^2+\bar c \chi\mathbb{W}_3
\mathbb{U}_3^* |\mathbb{U}_1|^2+ \sqrt{\frac{m_1}{m_3}}
\mathbb{W}_3^* \mathbb{W}_1 \mathbb{U}_1^* \mathbb{U}_3 \right)\nn\\
&=& \epsilon\, |Y_{\nu33} Y_{\nu 31} Y_{\nu21} Y_{\nu23}| \left[
-\frac{t_{23}}{c_{12}^3} \sin\frac{\alpha_2}{2} + {\cal
O}(s_{13}\sin\delta) \right]\,,
\eea
which is suppressed the more $\epsilon$ is small. Indeed, as already
mentioned, $I_4$ vanishes in the limit $m_1=0$ (namely
$\epsilon=0$), as a consequence of the alignment between the second
and third rows of $Y_\nu$, which ensures a vanishing eigenvalue.
The remaining contributions from $I^{(31),(32)}_e$ do not vanish
but, as stressed before, are negligible even considering future
sensitivities.

\end{itemize}

Summarising, it turns out that a large $d_e$ is obtained only under
the conditions of the first case, namely if $M_2\sim M_@$ and the
relation $m_1/m_@ \sim M_@/M_3 $ holds. The contribution induced by
$\delta$ could even exceed the one associated to $\alpha_2$; the
phase $\alpha_1$ plays no role. Notice in particular that values of
$m_1 \lesssim 0.5\times 10^{-3} m_3$ are incompatible with an
experimentally relevant $d_e$.

We remark that in all the expressions above, the Yukawa couplings,
the elements of the matrix $U$ and the light neutrino mass
eigenstates are consistently evaluated at $M_{\Pl}$. For instance,
in the case of NH, the effect of running is such that $\hat m$ is
rescaled by a numerical factor, while the parameters of $U$ do not
change significantly. We illustrate this point by considering a
numerical example for the first case discussed previously. If we
take the low energy neutrino spectrum consistent with the
present-day neutrino oscillation data, $\hat m=(0.96\times
10^{-2},0.17,1)m_@$, then, for $\tan\beta=30$, $(M_3,M_2)= (10^2,
2)M_@$, $M_1=3\times10^{11}$ GeV, $\alpha_2=\pi/2$, $\delta=\pi/2$
and any $\alpha_1$, at the Planckian scale we obtain\footnote{In
practice we do the opposite, namely we assign $U$ and rescale $\hat
m$  by a numerical factor at high energy and check that, when
evolved at low energy, the parameters of $U$ and the neutrino
spectrum are still within the experimental window.} $\hat m=1.8
(10^{-2},0.166,1)m_@$. Correspondingly, the mixing angles change
from $\theta_{23}=47^\circ$, $\theta_{12}=35.2^\circ$,
$\theta_{13}=9.7^\circ$ to $\theta_{23}=45^\circ,\theta_{12}=35^
\circ,\theta_{13}=10^\circ$. With these choices one obtains:
$I_e^{\FV}=10.2$, $I_e^{\FC}=1.35$, $C_{31}=1.5$, $C_{32}/8=
C_{21}=10^{-3}$. These values are slightly changed taking different
values of $\tan\beta$, which enters only through the running
effects. This means that, in the point $P$ and for any $\tan\beta$,
the present example gives $d_e^{\FC}=1.3\times 10^{-3} d_e^{exp}$,
which may escape detection. Since $d_e^{FV}$ and LFV decays
explicitly depend on $\tan\beta$, we need to specify it. Taking for
instance $\tan\beta=30$, one has $d_e^{\FV}=0.022 \,d_e^{exp}$,
which is at hand of future experiments. With $\tan\beta=30$, the BRs
of $\teg$, $\tmg$ and $\meg$ turn out to be smaller than their
corresponding experimental limits by factors of $0.8$, $4\times
10^{-5}$ and $0.02$, respectively.

\begin{figure}
\begin{center}
\begin{tabular}{cc}
\includegraphics[width=7.1cm]{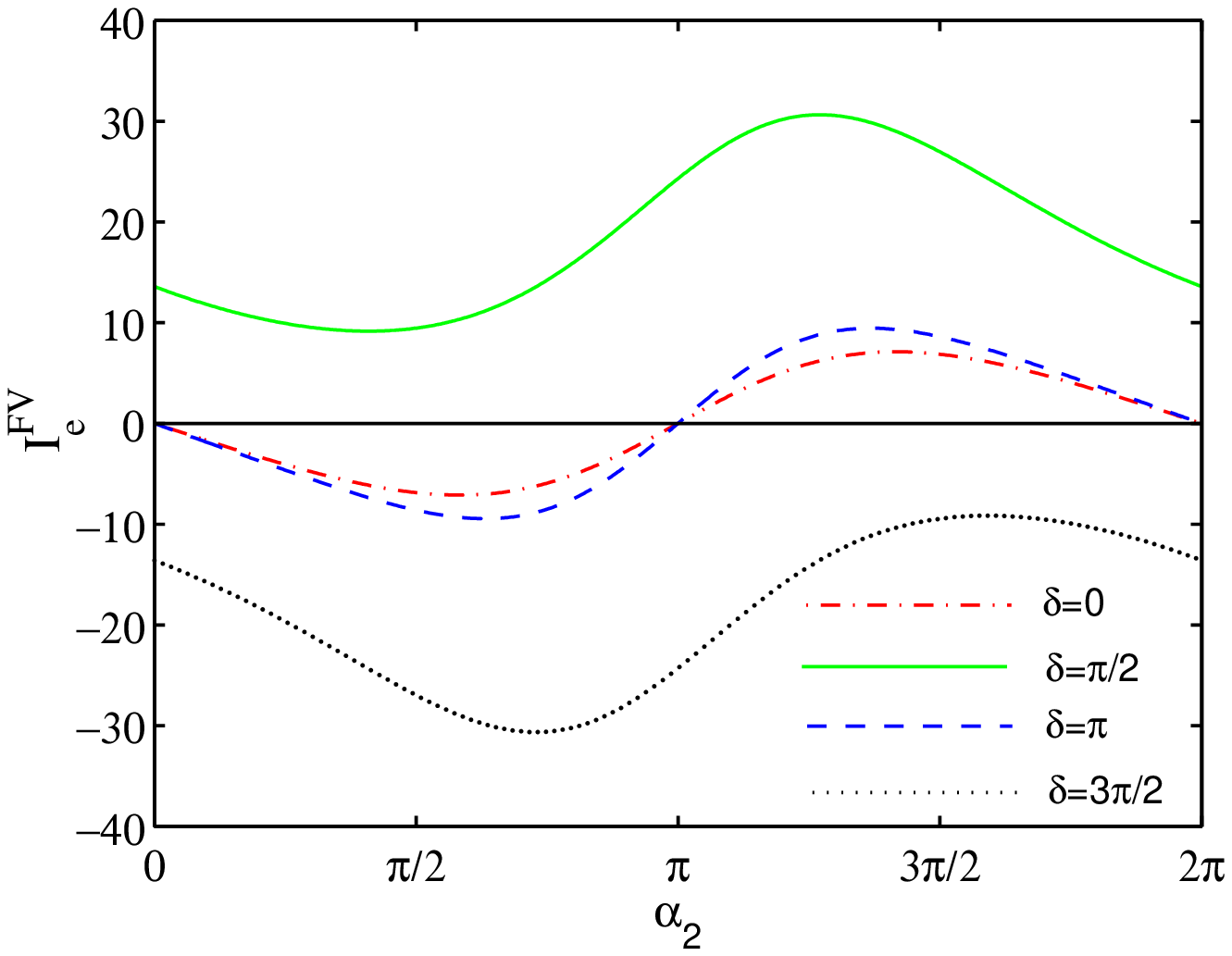}
&\includegraphics[width=7.1cm]{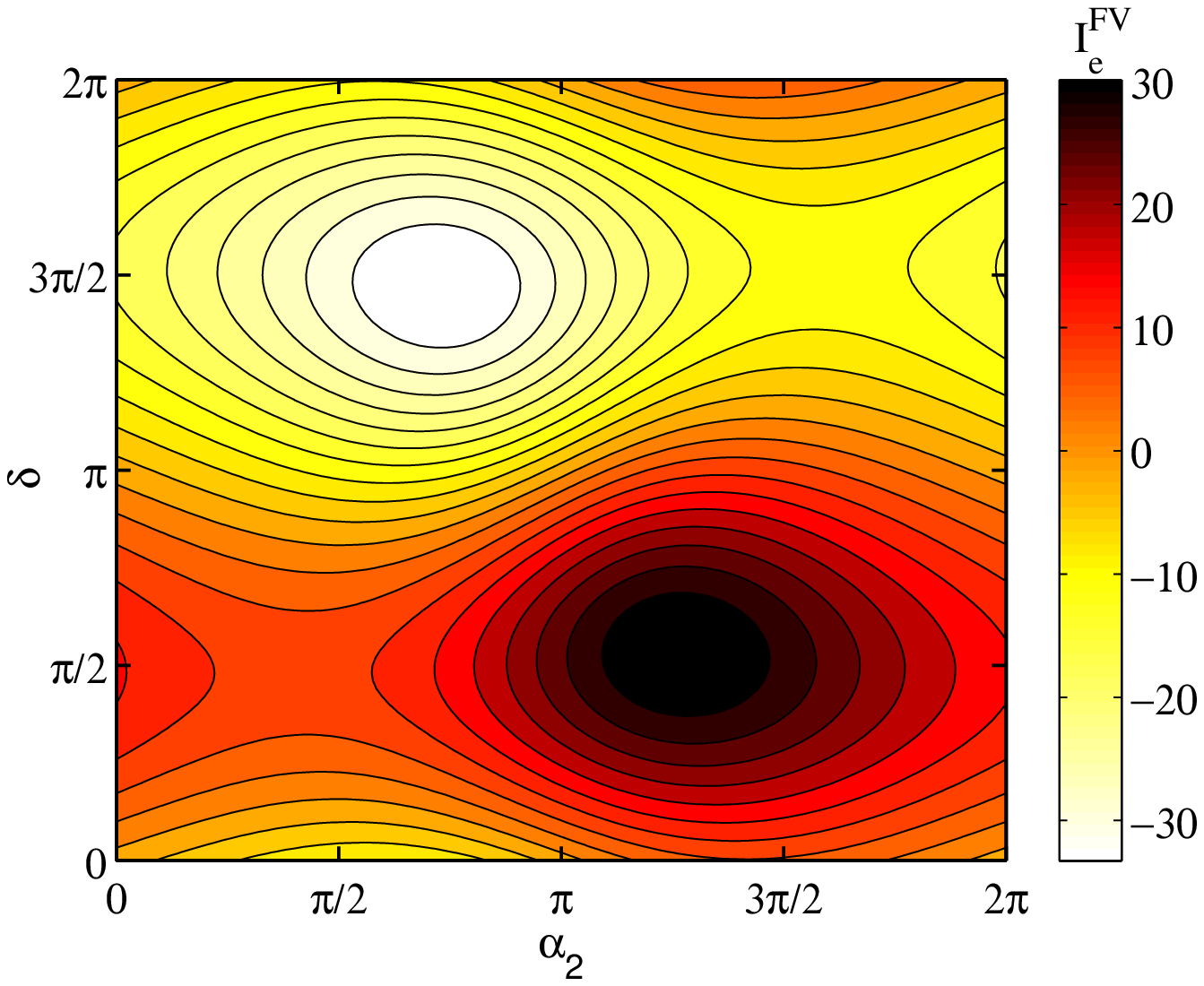}\\
\includegraphics[width=7.1cm]{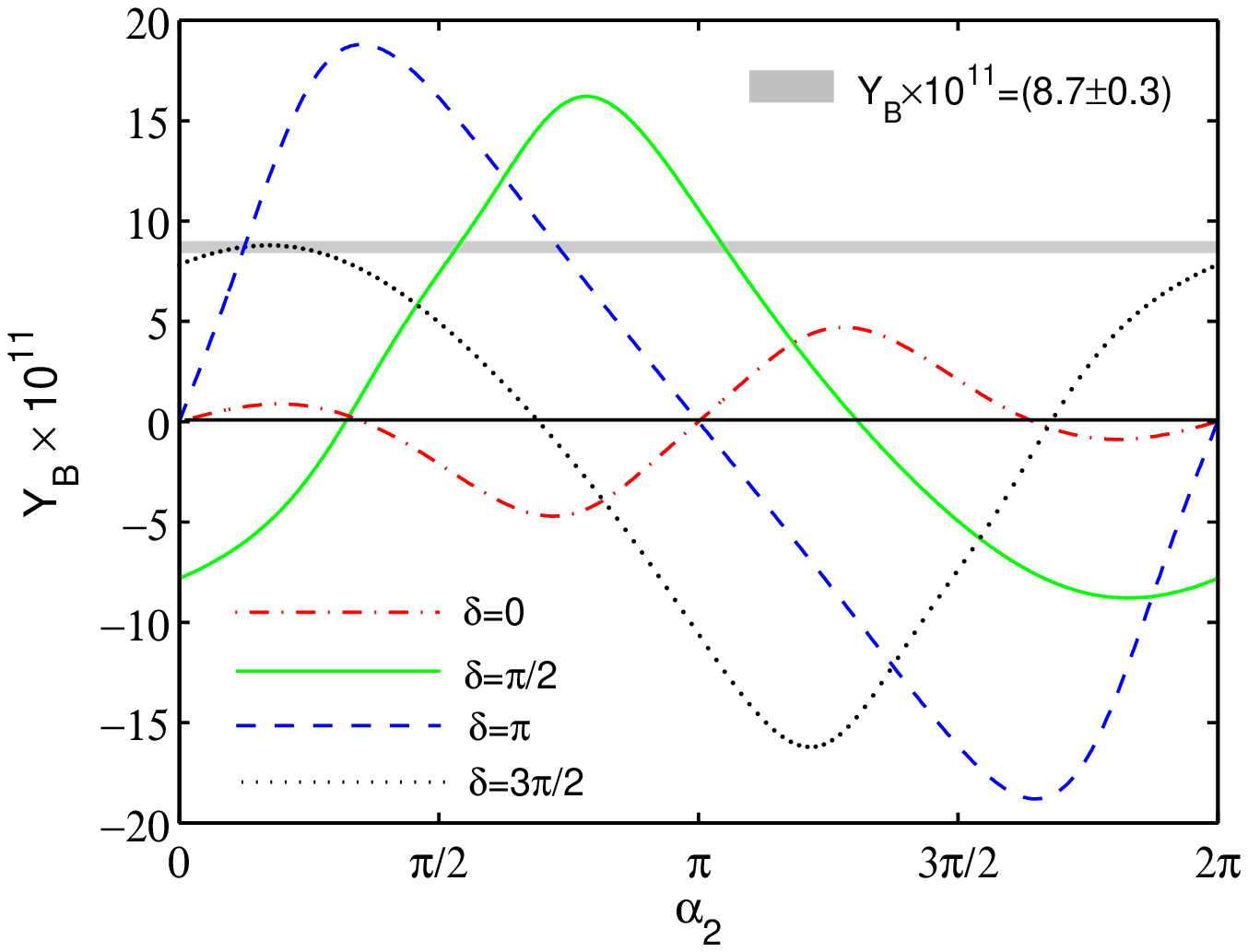}
&\includegraphics[width=7.1cm]{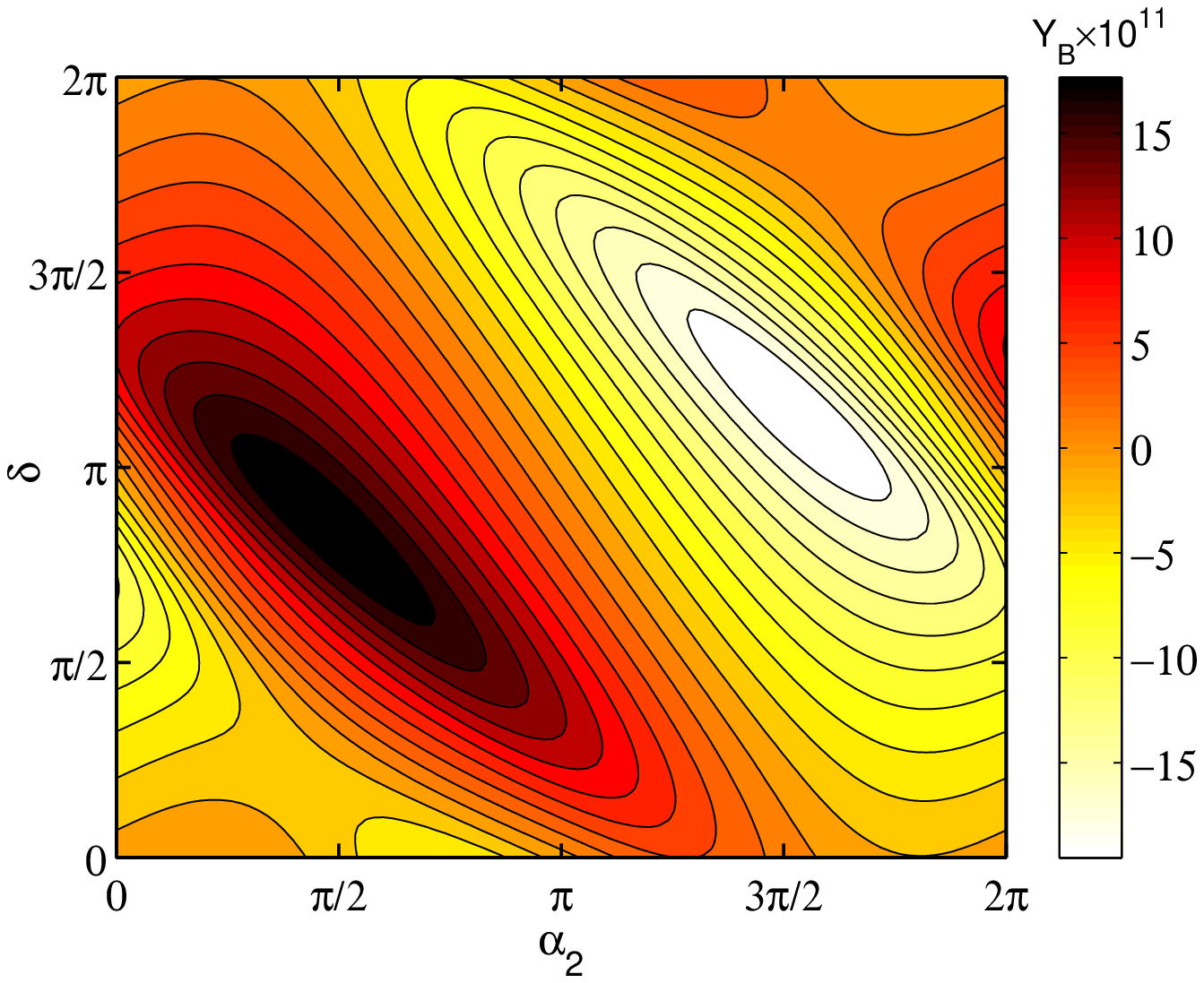}
\end{tabular}
\end{center}
\vspace*{-0.5cm} \caption{Dependence of $I_e^{\FV}$ (left
upper-plot) and $Y_B$ (left lower-plot) on $\alpha_2$ for a given
set of values for $\delta$ and $\tan\beta>10$ ($\mu$ and $\tau$ in
equilibrium): we take $\delta=0,\pi/2,\pi,3\pi/2$. On the right-hand
side we show the behaviour of $I_e^{\FV}$ (upper plot) and  $Y_B$
(lower-plot) in the $(\alpha_2,\delta)$-plane. For the choice of the
remaining parameters, see the text.}\label{figNH}
\end{figure}

\begin{figure}
\begin{center}
\begin{tabular}{cc}
\includegraphics[width=7.1cm]{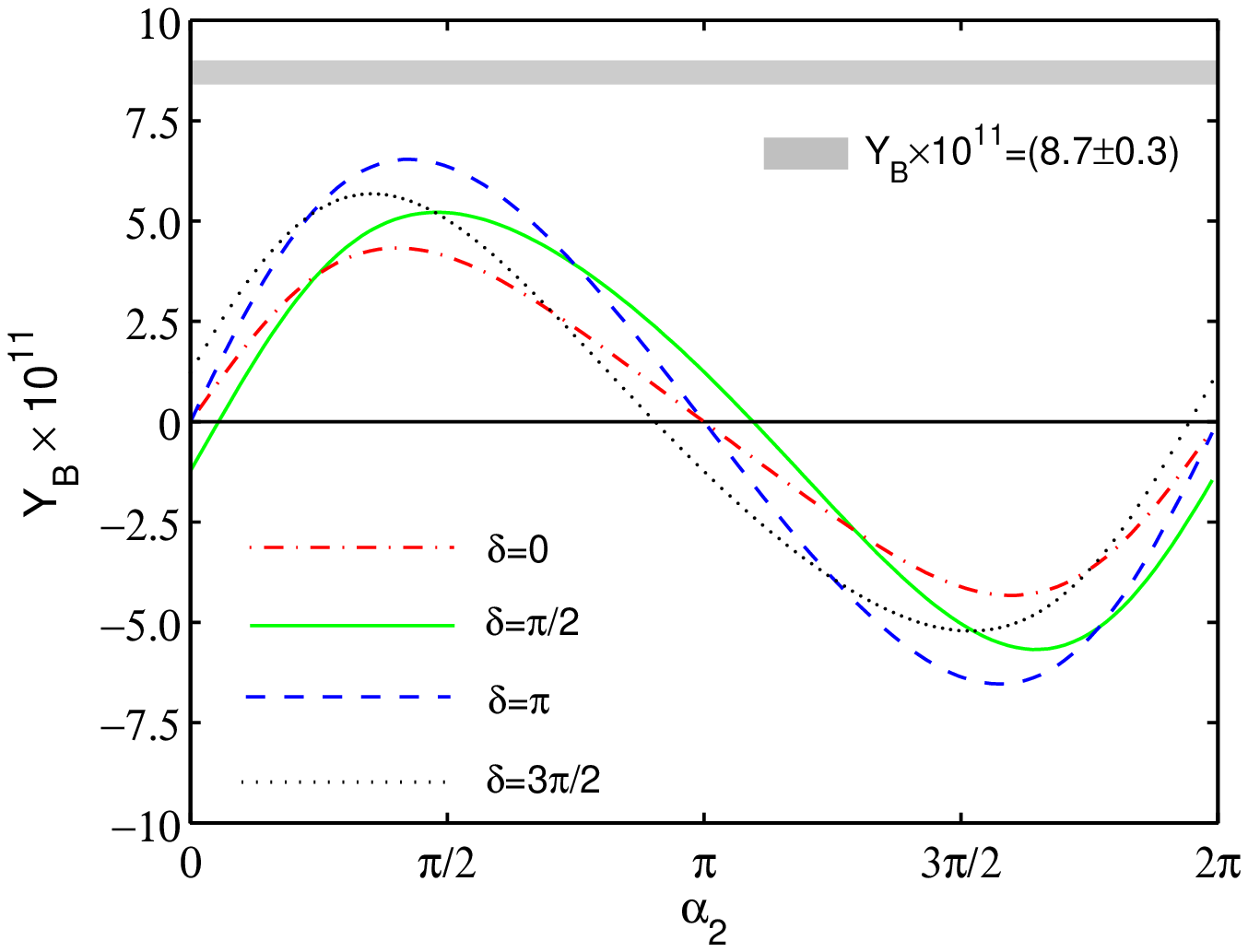}
&\includegraphics[width=7.1cm]{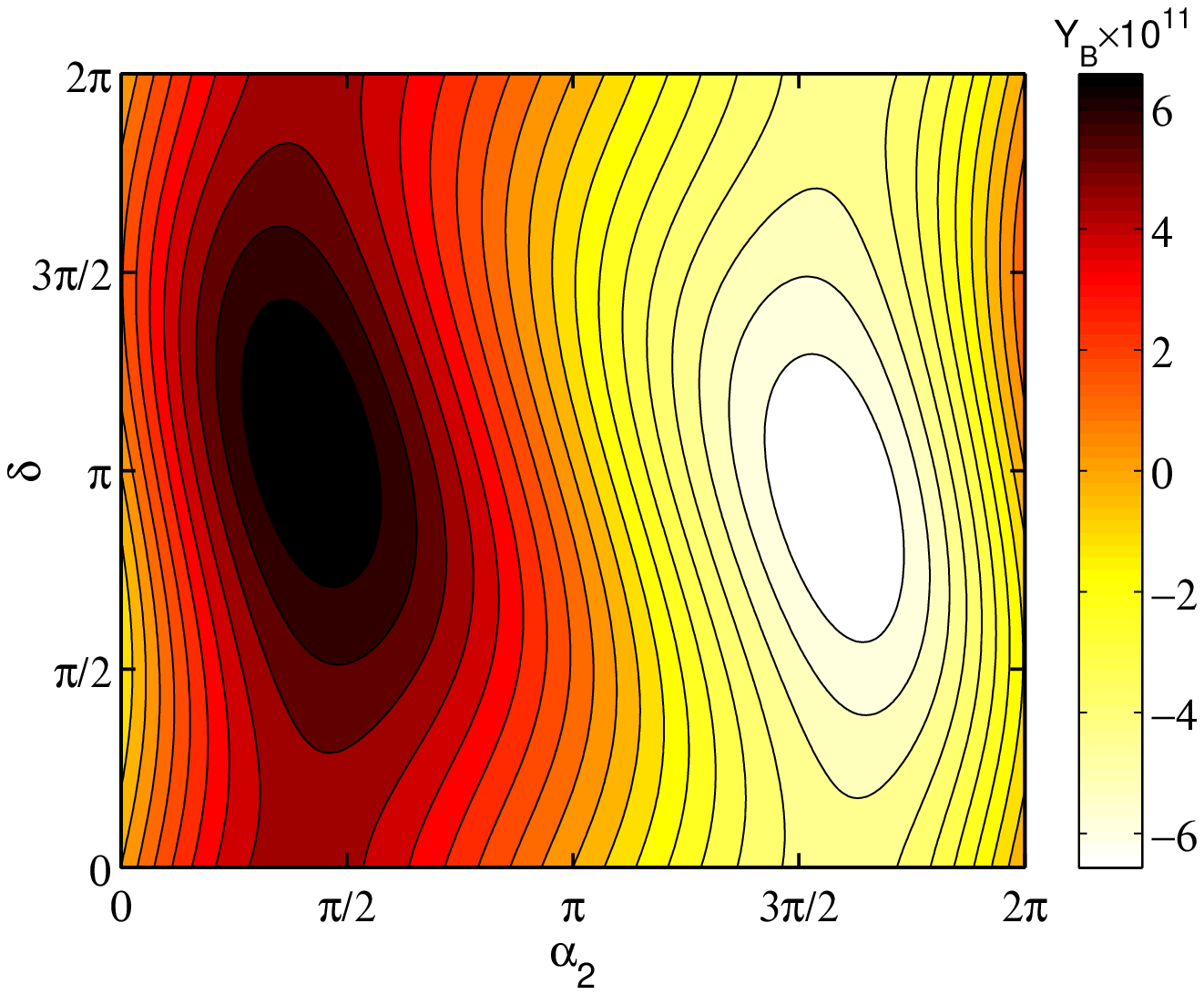}
\end{tabular}
\end{center}
\vspace*{-0.5cm} \caption{The same as in Fig.~\ref{figNH} for $Y_B$
taking $\tan\beta<10$ (only $\tau$ in equilibrium).}\label{figNH1}
\end{figure}

The left-hand side of Fig.~\ref{figNH} shows the dependence of
$I_e^{\FV}$ and $Y_B$ on $\alpha_2$ for a given set of values for
$\delta$, keeping for the remaining parameters of $U$ and light and
heavy neutrino masses the same set as before - in particular we
recall that $M_1=3\times 10^{11}$~GeV and $\tan\beta=30$, but the
dependence of $Y_\nu(M_{\Pl})$ on $\tan\beta$ is mild. On the
right-hand side of the same figure, we show the behaviour of
$I_e^{\FV}$ and $Y_B$ in the $(\alpha_2,\delta)$-plane. Hence,
$I_e^{\FV}$ displayed in Fig.~\ref{figNH} slightly changes
considering other values of $\tan\beta$. The same applies separately
for each asymmetry $\epsilon_\ell$, but not to $Y_B$, for which two
different regimes can be identified according to the value of
$\tan\beta$. Indeed, having $M_1=3\times 10^{11}$~GeV, in the case
that $\tan\beta> 10$ the baryon asymmetry is generated in the range
of temperatures for which all the lepton flavours, but the electron
one, are in thermal equilibrium and Eq.~(\ref{a}) applies, see
Fig.~\ref{figNH}; in the case that $\tan\beta<10$ thermal
leptogenesis takes place in a range of temperatures where only the
tau flavour is in thermal equilibrium and, in this case, one has to
apply Eq.~(\ref{aa}), see Fig.~\ref{figNH1}. From the comparison of
the upper and lower plots of Fig.~\ref{figNH} one can see that there
are regions in the $(\alpha_2,\delta)$ parameter space where the
seesaw-induced effects can lead to values of $d_e$ within future
experimental sensitivities and $Y_B$ is compatible with its observed
value. For $\tan\beta<10$ the baryon asymmetry is on the contrary
below its observed value (the horizontal grey line) and one would
need slightly larger values of the mass $M_1$ to obtain a value of
the baryon asymmetry consistent with observation. This seems to be a
generic conclusion. Also notice that, in this temperature regime,
$Y_B$ does not depend strongly on $\delta$.

Fig.~\ref{figmlight} shows the dependence of $I_e^{FV}$ and $Y_B$ on
$m_1/m_3$ (left and right panel respectively), keeping $m_1/m_3=M_@/M_3$ as in the previous example,
as well as the same set for the remaining parameters.
We also display (dashed line) the baryon asymmetry computed
in the one-single flavour approximation.
We see that there is almost one order of magnitude difference.

%%%%%%%%%%%%%%%%%%%%%%%%%%%%%%%%%%%%%%%%%%%%%%%%%%%%%%%%%%%%%%%%%%

\subsection{Inverted hierarchy for light neutrinos}

For inverted hierarchy (IH), we take $m_{2,1}\approx  m_@$ and leave
$m_3$ undetermined. It is convenient to redefine the matrix $R$
according to: $R=O_{12}O_{13}O_{23}\times{\rm antidiag}(1,1,1)$.
This simple exchange of the first and third columns of $R$ allows to
carry on the discussion in a parallel way to what done before.
Indeed, assuming also $R_{i1}\neq 0$, the conditions (\ref{la1})
become
\beq
-\frac{R_{i2}}{R_{i1}}= \widetilde t + \frac{R_{i3}}{R_{i1}} \,\bar
t\,,%
\label{eqRIH}
\eeq
where $\widetilde t$ and $\bar t$ are defined as in Eq.~(\ref{tt}),
but now we have $|\widetilde t|\approx 1/\sqrt{2}$ while $|\bar
t|\ll 1$.

Using these expressions, the Yukawa couplings of the second and
third rows are given by
\bea
Y_{\nu2i}=\sqrt{\frac{M_2}{M_@}} \left(R_{21}{\mathbb W_i^*}  +
R_{23}  \sqrt{\frac{m_3}{m_1}} \mathbb{U}^*_i \right)\,,\nn\\
Y_{\nu3i}=\sqrt{\frac{M_3}{M_@}} \left(R_{31}  {\mathbb W_i^*}  +
R_{33} \sqrt{\frac{m_3}{m_1}} \mathbb{U}^*_i \right)\,,
\eea
which, as before, correctly vanish for $i=2$. Large $Y_{\nu21}$
requires $M_2 ={\cal O}(M_@)$, so that to maximize $d_e$ we require
$M_3\gg M_@$. Instead, large $Y_{\nu 31}$ is obtained provided
$R_{13} ={\cal O}(\sqrt{M_@/M_3}) < 1$. Explicitly, $R_{31}=c^R_{23}
c^R_{13}$, but the condition for $i=3$ of Eq.~(\ref{eqRIH}),
$t^R_{23}= \widetilde t - t^R_{13} \bar t/c^R_{23}$, suggests that
$c^R_{23}$ is quite large. We rather suppress $c^R_{13} =\chi <1$
and expand at first order in $\chi$. In particular, now
$c^R_{23}\,\chi\approx\sqrt{M_@/M_3}Y_{\nu31}/\mathbb{W}_1^*$ and
the conditions in (\ref{eqRIH}) get simplified:
\beq
\frac{c^R_{12} c^R_{23} + s^R_{12} s^R_{23}}{-c^R_{12} s^R_{23} +
s^R_{12} c^R_{23}}+{\cal O}(\chi\bar t) = \widetilde t = t^R_{23} +
\frac{1}{\chi c^R_{23}} \bar t\,.
\label{simplyIH}%
\eeq
It is useful to distinguish between two relevant cases.

\begin{itemize}

\item If $s^R_{12}=\epsilon <1$, Eq.~(\ref{simplyIH}) gives $t^R_{23}
=-1/\widetilde t +\epsilon$ and $\bar t \approx (\widetilde
t+1/\widetilde t) \chi c^R_{23}$. The latter can be rewritten as
\beq
\sqrt{\frac{m_3}{m_2}} \sqrt{\frac{M_3}{M_@}} ={\cal O}(1)\,. %Y_{\nu31}
\eeq
Notice that in this case $m_3$ cannot be arbitrarily small, otherwise
$M_3$ would exceed the GUT or Planck scale.
%$s^R_{23}\approx-\tilde c$, $c^R_{23}\approx\tilde s$,
In addition
\beq
R = \left( \matrix{ \widetilde s  & \widetilde c  & \chi \cr
-\widetilde c  &  \widetilde s &  0 \cr \widetilde s \chi & \chi
\widetilde c & -1 } \right) +{\cal
O}(\epsilon^2,\epsilon\chi,\chi^2)\,.
\eeq
The large masses $m_2$ and $m_1$ are dominated by $N_1$ and $N_2$
with competitive strenght\footnote{Assuming for instance
$\alpha_1=\alpha_2$ and tri-bimaximal mixing one has $\widetilde c =
\sqrt{2/3}$, $\widetilde s = \sqrt{1/3}$; while for
$\alpha_2-\alpha_1=\pi$ one has $\widetilde c = \sqrt{2}$,
$\widetilde s =-i$.}. In particular, the first row of $R$ determines
leptogenesis, as we are going to discuss. The
smallest mass $m_3$ is instead dominated by $N_3$, which decouples
the more $\chi$ is small. The Yukawas are
\beq
Y_{\nu3i}=    \sqrt{\frac{M_3}{M_@ }} \left(\widetilde s \chi
\mathbb{W}^*_i - \sqrt{\frac{m_3}{m_1}}  \mathbb{U}^*_i \right)~~,~~
Y_{\nu2i}=   \sqrt{\frac{M_2}{M_@}} \left(-\widetilde c
\mathbb{W}^*_i +{\cal O}(\chi \epsilon) \sqrt{\frac{m_3}{m_1}}
\mathbb{U}^*_i\right)\,,
\eeq
so that
\beq
I_4 =  \frac{M_2}{M_@} \frac{M_3}{M_@} |\widetilde c|^2
\sqrt{\frac{m_3}{m_1}} \,\,{\rm Im}\!\left( -\widetilde s \chi
\mathbb{U}_1 \mathbb{W}_1^* |\mathbb{W}_3|^2+\widetilde s
\chi\mathbb{U}_3 \mathbb{W}_3^* |\mathbb{W}_1|^2 +
\sqrt{\frac{m_3}{m_1}} \mathbb{U}_3^* \mathbb{U}_1 \mathbb{W}_1^*
\mathbb{W}_3 \right)\,,\nn\\
\label{i4}
\eeq
which turns out to be proportional to $U_{e3}$.

\item  If $c^R_{12}=\epsilon <1$, one has $t^R_{23} = \widetilde t\,
(1-\epsilon\widetilde t)$ and $\bar t \approx \epsilon  \,\widetilde
t\,(\widetilde t+1/\widetilde t\,) \chi c^R_{23}$, which in turn
implies
\beq
\sqrt{\frac{m_3}{m_2}}  \sqrt{\frac{M_3}{M_@}} ={\cal O}( \epsilon)
\eeq
In this case $m_3$ is allowed to be small and the limit
$m_3\rightarrow 0$ can be applied. In addition
\beq
R= \left( \matrix{ \widetilde s &   \widetilde c  &   0 \cr
-\widetilde c &  \widetilde s &  - \chi \cr \widetilde c \chi  &
-\chi \widetilde s  & -1 } \right) +{\cal
O}(\epsilon^2,\epsilon\chi,\chi^2)\,.
\eeq
The Yukawas are
\beq
Y_{\nu3i}=    \sqrt{\frac{M_3}{M_@}} \left(\widetilde c \chi
\mathbb{W}^*_i - \sqrt{\frac{m_3}{m_1}}  \mathbb{U}^*_i \right)~~,~~
Y_{\nu2i}=    \sqrt{\frac{M_2}{M_@}} \left(-\widetilde c
\mathbb{W}^*_i -{\cal O}(\chi)\sqrt{\frac{m_3}{m_1}} \mathbb{U}^*_i
\right)\,,
\eeq
so that
\beq
I_4 =  \frac{M_2}{M_@} \frac{M_3}{M_@} |\widetilde c|^2
\sqrt{\frac{m_3}{m_1}} {\rm Im}\left( -\widetilde c \chi
\mathbb{U}_1 \mathbb{W}_1^* |\mathbb{W}_3|^2+\widetilde c
\chi\mathbb{U}_3 \mathbb{W}_3^* |\mathbb{W}_1|^2
+ \sqrt{\frac{m_3}{m_1}} \mathbb{U}_3^* \mathbb{U}_1
\mathbb{W}_1^* \mathbb{W}_3 \right)\,,\nn\\
\eeq
which is again proportional to $U_{e3}$ and suppressed by a factor
of $\epsilon$ with respect to (\ref{i4}). Hence, it vanishes in the
limit $m_3=0$ due to the fact that the second and third rows of
$Y_\nu$ are aligned. The contributions from $I^{(31),(32)}_e$ are
negligible even for the future sensitivities.

\end{itemize}

Summarizing, in the inverted-hierarchical case $d_e$ can be at hand
of future experiments only if $M_2\sim M_@$, $m_3/m_@ \sim M_@/M_3$
and $|U_{e3}|$ is large. The dependence on $\alpha_2$ and $\delta$
is quite complicated and also $\alpha_1$ plays a role.

With an IH light neutrino spectrum, the RGE effects might be
important, in particular for the solar angle and mass squared
difference. Nevertheless, the above conclusions remain valid. In
practice, one has just to rescale $\hat m$ and $m_\odot$ at high
energy by a numerical factor, and check whether the parameters of
$U$, when evolved at low energy, fall within the experimental
window. It is well known~\cite{RGEIH} that this is always the case
if $|\alpha_2-\alpha_1|=\pi$, namely if the solar pair of
eigenstates are of pseudo-Dirac type. Then, $I_e^{\FV}$ and all the
$C_{ij}$ are mildly dependent on $\alpha_2$ and essentially depend
only on $\delta$. This will turn out to be the case also for $Y_B$.

We illustrate this considering an example of the first case where
the RGE effects turn out to be quite small. We consider
$\tan\beta=30$ and take at $M_{\Pl}$: $\hat m=1.5
(1,0.97,10^{-2})m_@$, $(M_3,M_2)= (8, 0.1)M_@$, $M_1=10^{11}$ GeV,
$\theta_{23}=45^\circ,\theta_{12}=35^\circ,\theta_{13}=10^\circ$,
$\delta=3\pi/2$, and any $\alpha_2=\alpha_1+\pi$. At low energy, the
neutrino spectrum is viable, $\hat m=(1,0.98,10^{-2})m_@$, and the
angles of the MNS are $\theta_{23}=44.4^\circ$,
$\theta_{12}=35^\circ$, $\theta_{13}=10.4^\circ$. We obtain
$I_e^{\FV}=-4.8$, $I_e^{\FC}=-0.45$, $C_{31}=1.4$, $C_{32}/4=
C_{21}=3\times 10^{-3}$. Hence, for the point $P$,
$d_e^{\FC}=-5\times 10^{-4} d_e^{exp}$. With $\tan\beta=30$, the FV
contribution to $d_e$ is at hand of future experiments
$d_e^{\FV}=0.01 d_e^{exp}$; as for the BR of $\teg$, $\tmg$, $\meg$,
they are smaller than their corresponding experimental limits by
factors of $0.7$, $4\times 10^{-7}$, $0.15$, respectively.

\begin{figure}
\begin{tabular}{cc}
\includegraphics[width=7.6cm]{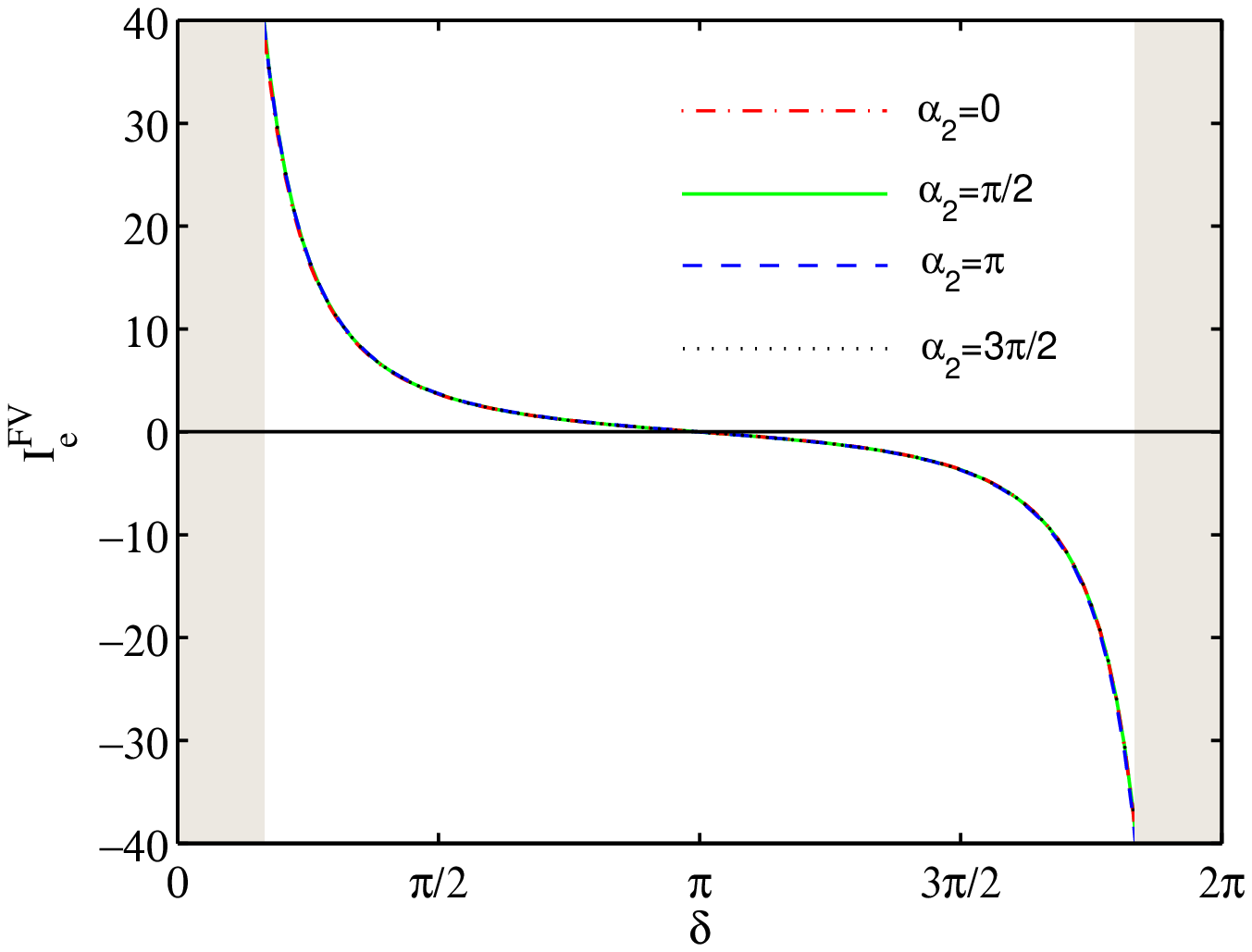}
&\includegraphics[width=7.6cm]{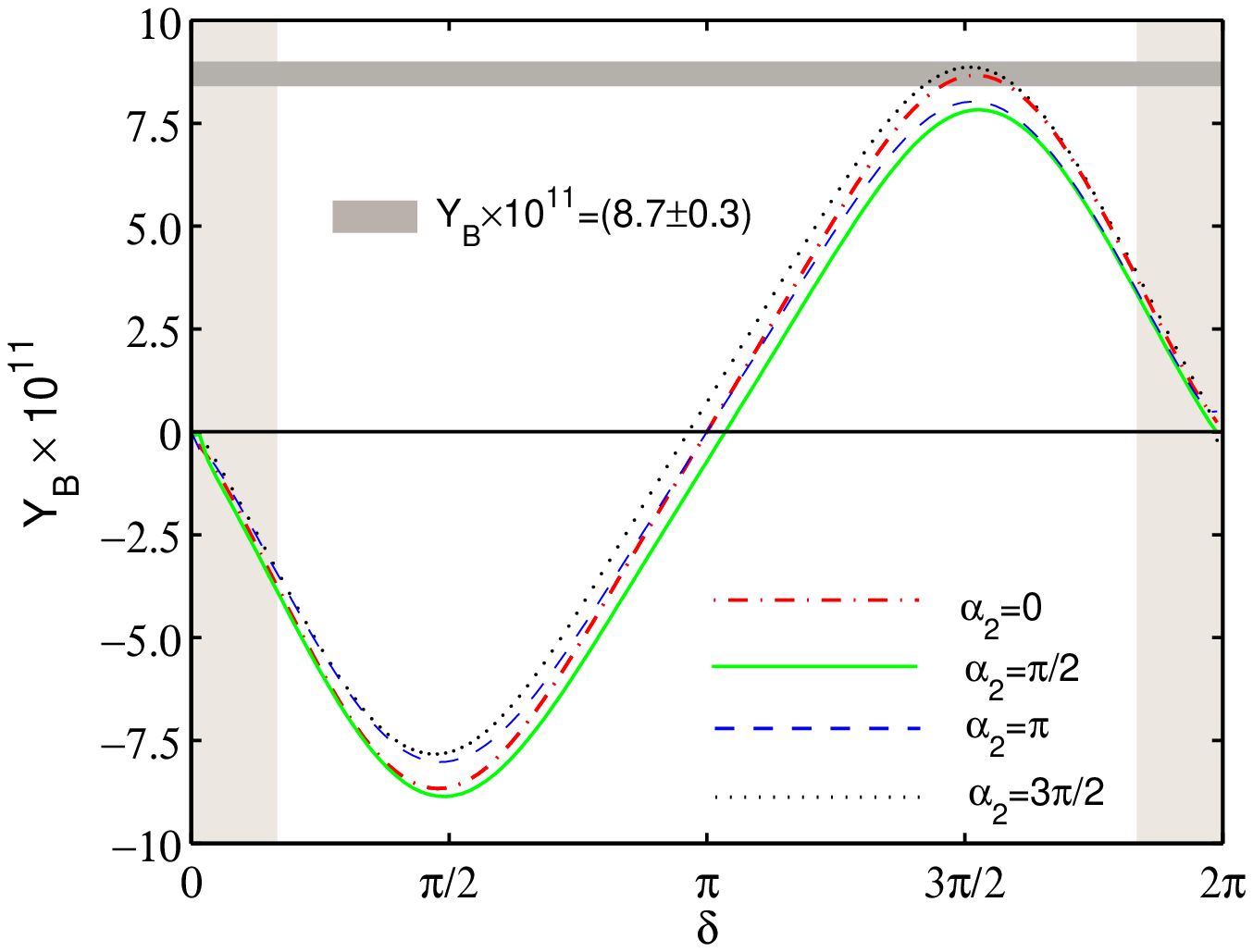}
\end{tabular}
\caption{Dependence of $I_e^{\FV}$ and $Y_B$ on $\delta$ for
$\alpha_2=\alpha_1+\pi$: we take $\alpha_2=0,\pi/2,\pi,3\pi/2$. The
vertical light-grey regions are excluded by requiring perturbative
$Y_\nu$. For the choice of the other parameters, see the text.}
\label{figIH}
\end{figure}

Fig.~\ref{figIH} shows the dependence of $I_e^{FV}$ and $Y_B$ on
$\delta$, keeping for the remaining parameters the same set as
before. The curves are absent for $\delta$ very small or close to
$2\pi$, where $C_{31}>10$ because some of the Yukawas in $Y_\nu$
blow up. Since both $Y_B$ and $I_e^{\FV}$ are directly proportional
to $|U_{e3}|$, the plot can be adapted correspondingly to other
values of $\theta_{13}$. For a sizeable $d_e$ and $Y_B$,
$\theta_{13}$ cannot be smaller than a few degrees. The value chosen
for $M_1$ is $10^{11}$ GeV which corresponds to the regime of
temperatures where all lepton flavours, but the electron one, are in
thermal equilibrium, since $\tan\beta=30$. In this case,
Eq.~(\ref{a}) applies. We have checked that for smaller values of
$\tan\beta$, where only the tau is in equilibrium, the final baryon
asymmetry does not change significantly.

\begin{figure}
\begin{center}
\includegraphics[width=15cm]{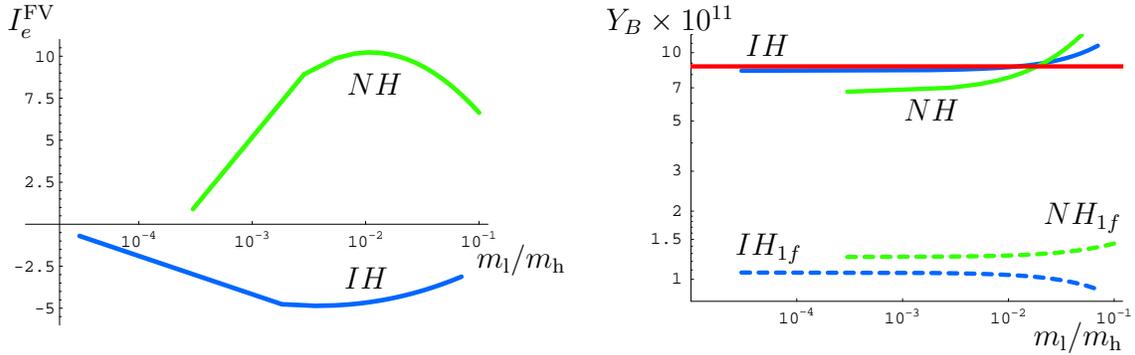}
\end{center}
\vspace*{-0.5cm} \caption{Dependence of $I_e^{\FV}$ and $Y_B$ on
$m_{\rm{l}}/m_{\rm{h}}$. For normal hierarchy
$m_{\rm{l}}/m_{\rm{h}}=m_{\rm{1}}/m_{\rm{3}}$, while inverted
hierarchy $m_{\rm{l}}/m_{\rm{h}}=m_{\rm{3}}/m_{\rm{2}}$. For the
choice of the other parameters, see the text. The dashed lines are
the result of the 1-flavour approximation.}\label{figmlight}
\end{figure}

In Fig. \ref{figmlight} we show the dependence of $I_e^{\FV}$ and
$Y_B$ on $m_3/m_2$ (left and right panels, respectively), while
keeping $m_3/m_2 = 0.08\,M_@/M_3$ as in the previous example and
assuming the same set for the remaining parameters (in particular we
selected $\alpha_2=0$, $\alpha_1=\pi$). On the right-hand side we
compare the baryon asymmetry computed with the flavour effects
included to the baryon asymmetry computed in the one-single flavour
approximation. We see that there is almost one order of magnitude
difference, both in the normal and in the inverted hierarchical
case. The main reason is that in the one-flavour case the total
lepton asymmetry is strongly washed out since the wash-out parameter
$\widetilde{m}=\widetilde{m}_e+\widetilde{m}_\mu
+\widetilde{m}_\tau$ is larger than $m_@$, yielding a suppressed
baryon asymmetry. On the contrary, in the flavour approximation, the
asymmetries in the electron and muon flavours are only weakly washed
out. Furthermore, in the inverted hierarchy case, the one-flavour
approximation leads to a suppression in the CP asymmetry that goes
like $m_\odot^2/m_@$, while including flavours the individual CP
asymmetries go as $m_@$.

\section{Conclusions}
\label{conclusions}

The predictions for the electron EDM in the context of the
supersymmetric seesaw and mSUGRA have been investigated. First, we
showed the existence of an indirect upper bound on $d_e$ from
negative searches for $\teg$, within the supersymmetric seesaw.
This indirect bound may be even stronger than the present experimental direct upper limit.
Planned searches for $d_e$, improving the sensitivity by about three
orders of magnitude, will be able to supersede the indirect bound
from $\teg$ and provide considerable tests of the seesaw-induced
effects. %Requiring $M_1\lesssim {\cal O}(10^{11})$ GeV,
We identified in a model-independent way the patterns of seesaw
models that lead to a potentially observable electron EDM,
considering in turn the case of normal and inverted light neutrino
spectra. A widely splitted spectrum for right-handed neutrinos is a
crucial ingredient, as well as the relation $m_{\rm l}/m_{\rm h}
\sim M_@/M_{3}$, where $M_@ = v_u^2/m_@$. Indeed, in the limit of a
vanishing lightest neutrino mass, the electron EDM drops much below
the planned sensitivities.

The seesaw interactions may also be responsible for the generation
of the baryon asymmetry of the Universe via the mechanism of
leptogenesis. The patterns of seesaw models identified requiring an
electron EDM within future experimental sensitivities allow to
extract the value of $M_1$ required to generate a sufficiently large
baryon asymmetry via thermal leptogenesis. The importance of taking
into account flavour effects has been emphasized.  Our findings show
that a large enough baryon asymmetry may be achieved through thermal
leptogenesis for those patterns which give rise to a large electric
EDM and suppressed $\teg$, $\tmg$, $\meg$ branching ratios. However,
a sufficiently large baryon asymmetry is reached only for large
values of $M_1$, at least larger than about $10^{11}$ GeV. Since we
are dealing with a  supersymmetric leptogenesis set-up, we should
face the problem arising from the so-called gravitino bound. The
latter is posed by the possible overproduction of gravitinos during
the reheating stage after inflation, see for instance~\cite{grav}.
Being only gravitationally coupled to the SM particles, gravitinos
may decay very late jeopardising the successfull predictions of Big
Bang nucleosynthesis. This does not happen, however, if gravitinos
are not efficiently generated during reheating, that is if the
reheating temperature $T_{RH}$ is bounded from above,
$T_{RH}\lesssim 10^{10}$~GeV~\cite{grav}. The severe bound on the
reheating temperature makes the generation of the RH neutrinos
problematic (for complete studies see~\cite{lept,aat}), if the
latter are a few times heavier than the reheating temperature,
rendering the thermal leptogenesis scenario unviable.

In view of the above, if a large $d_e$ is measured in the near
future and assuming that its main contribution comes from
CP-violating effects induced by the see-saw Yukawas, either the
baryon asymmetry is not explained within the thermal leptogenesis
scenario or leptogenesis occurs in a non-thermal way. This second
alternative stems from the fact that the RH neutrinos might be
generated not through thermal scatterings, but by other mechanisms,
for example during  the preheating stage~\cite{preh}, from the
inflaton decays~\cite{preh,asaka} or quantum fluctuations~\cite{gr}.
In these cases, the baryon asymmetry depends crucially on the
abundance of RH neutrinos and sneutrinos generated non-thermally.
For instance, these heavy states may be produced very efficiently
during the first oscillations of the inflaton field during
preheating up to masses of order $(10^{17}-10^{18})$
GeV~\cite{preh}; the final baryon asymmetry is generated of the
right-order of magnitude if the flavour lepton asymmetries satisfy
the mild condition $\epsilon_\ell \gtrsim 10^{-8}(10^{10}\,{\rm
GeV}/ T_{RH})(M_1/10^{11}\,{\rm GeV})$, as one can readily deduce
from Ref.~\cite{preh}.

\section*{Acknowledgements} F.R.J. thanks A. Rossi for many enlightening
discussions. A.R. thanks the CERN Theory Group where part of this work was done.
The work of F.R.J. is supported by {\em
Funda\c{c}\~{a}o para a Ci\^{e}ncia e a Tecnologia} (FCT, Portugal)
under the grant \mbox{SFRH/BPD/14473/2003}, INFN and PRIN Fisica
Astroparticellare (MIUR). We also acknowledge the EC RTN Network
MRTN-CT-2004-503369.

%%%%%%%%%%%%%%%%%%%%%%%%%%%%%%%%%%%%%%%%%%%%%%%%%%%%%%%%%%%%%%%%%%%%%%%%%%%%%
\newpage

\appendix

\section{Constraints from LFV e EDM}

\begin{figure}[h!]
\begin{center}
\includegraphics[width=12cm]{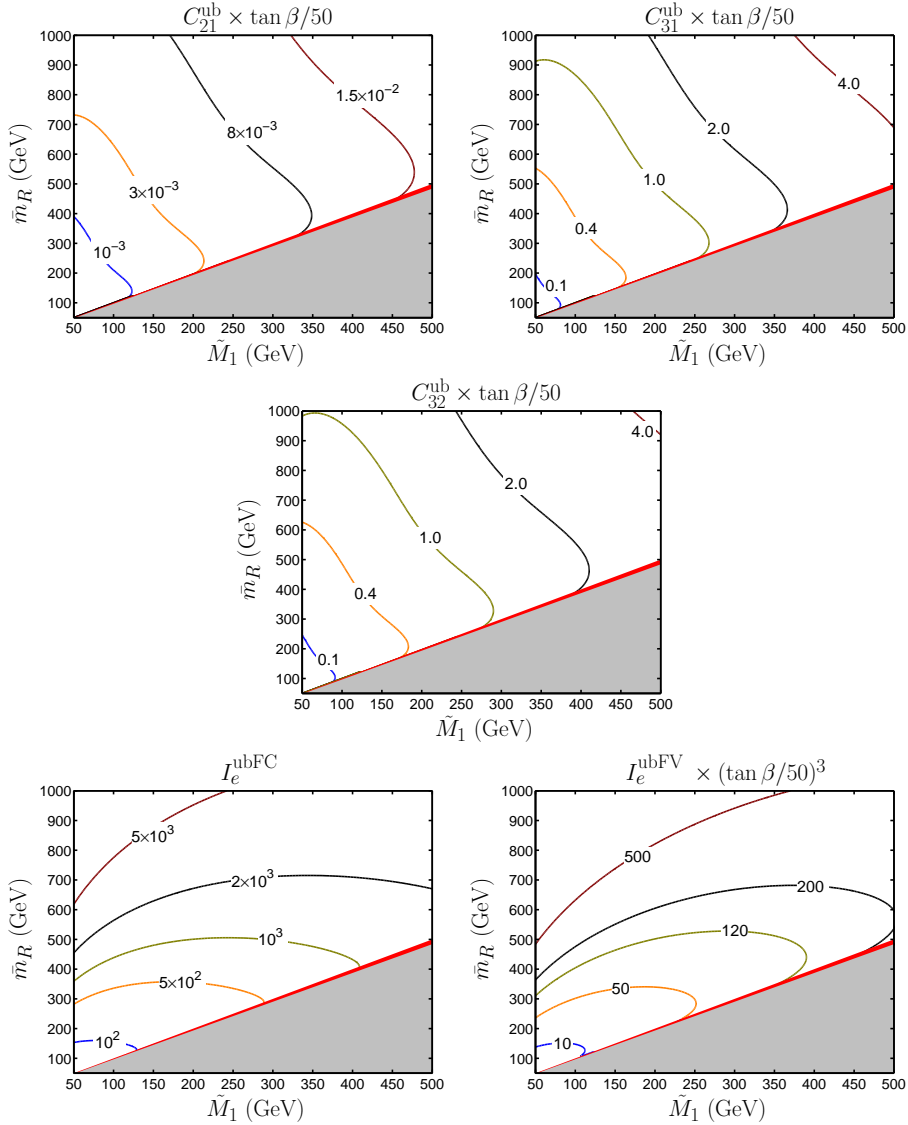}
\end{center}
\vspace*{-0.5cm}%
\caption{Contours of $C_{ij}^{\rm ub}$ (upper plots), from the
present $90 \%$ C.L. limits~\cite{PDG2006}. The lower plots show the
contours of $I_e^{\rm ubFC}$ and $I_e^{\rm ubFV}$ taking $d_e < 1.6
\times 10^{-27}$ e cm.} \label{fig1}
\end{figure}

In Fig.~\ref{fig1} we display $C_{ij}^{\rm ub}$, $I_e^{\rm ubFC}$
and $I_e^{\rm ubFV}$ from the present $90 \%$ C.L. limits~\cite{PDG2006},
${\rm BR}(\meg)< 1.2\times 10^{-11}\;,\; {\rm BR}(\tmg)<6.8\times
10^{-8} \;,\; {\rm BR}(\teg)< 1.1\times10^{-7}$.
The upper bounds $I_e^{\rm ubFC}$ and $I_e^{\rm ubFV}$ are also
shown taking $d_e < 1.6 \times 10^{-27}$ e cm.
We consider mSugra with $a_0=m_0+M_{1/2}$. For better sensitivities,
the values of $C^{\rm ub}_{ij}$ have to be multiplied by a factor
$\sqrt{{\rm BR}^{\rm fut}/{\rm BR}^{\rm pr}}$, while those of
$I_e^{\rm FCub}$ and $I_e^{\rm FVub}$ by a factor $d_e^{\rm
fut}/d_e^{\rm pr}$. We recall that $\tilde M_1 \simeq 0.4  M_{1/2}$
and $\bar m_R^2 \simeq m_0^2 + 0.15 M_{1/2}^2$. The plots are
adapted from those in~\cite{Masina:2005am}, where the reader can
find more details and references to the literature.

\newpage

%%%%%%%%%%%%%%  Bibliography   %%%%%%%%%%%%%%%%%%%%%%%%%%%%%%%

%%%%%%%%%%%%%%%%%%%%%%%%%%%%%%%%%%%%%%%%%%%%%%%%%%%%%%%%%%%
\end{document}